\newcommand*{\ATLASLATEXPATH}{latex/}
\documentclass[cernpreprint,texlive=2011,txfonts,UKenglish]{latex/atlasdoc} 
\usepackage{\ATLASLATEXPATH atlaspackage}
\usepackage{multirow}
\usepackage{longtable}
\usepackage{siunitx}
\usepackage{\ATLASLATEXPATH atlascontribute}
\usepackage[BSM=true]{\ATLASLATEXPATH atlasphysics}
\graphicspath{{logos/}{figs/}}

\usepackage{lfvdefs}
\usepackage{comment}
\usepackage{caption}
\DeclareCaptionStyle{capstyle}[justification=centering]{format=hang, labelfont={rm, bf}, textfont={rm} }
\AtlasTitle{Search for new phenomena in different-flavour high-mass dilepton
final states in \boldmath{$pp$} collisions at \boldmath{$\sqrt{s}=13$}
\si{\TeV} with the ATLAS detector}
\usepackage{authblk}
\author{The ATLAS Collaboration}

\date{\today}

\AtlasRefCode{EXOT-2015-20}
\PreprintIdNumber{CERN-EP-2016-168}
\AtlasJournal{EPJC}
\AtlasJournalRef{Eur. Phys. J. C76 (2016) 541}
\AtlasDOI{10.1140/epjc/s10052-016-4385-1}

\AtlasAbstract{%
A search is performed for a heavy particle decaying into different flavour
dilepton pairs ($e\mu$, $e\tau$ or $\mu\tau$), using \SI{3.2}{\per\femto\barn}
of proton--proton collision data at $\sqrt{s}=13$~\si{\TeV} collected in 2015
by the ATLAS detector at the Large Hadron Collider. No excess over the Standard
Model prediction is observed. Limits at the 95\% credibility level are set on
the mass of a \Zprime boson with lepton-flavour-violating couplings at 3.0,
2.7 and 2.6~\si{\TeV}, and on the mass of a supersymmetric $\tau$ sneutrino
with $R$-parity-violating couplings at 2.3, 2.2 and 1.9~\si{\TeV}, for
$e\mu$, $e\tau$ and $\mu\tau$ final states, respectively. The results are also
interpreted as limits on the threshold mass for quantum black hole production.}

\AtlasCoverSupportingNote{Supporting documentation}{https://cds.cern.ch/record/2135369}

\AtlasCoverCommentsDeadline{-}

\AtlasCoverAnalysisTeam{C. Blocker, M. Cano Bret, W. S. Chan, L. Flores,
D. Gingrich, J. Guo, L. Han, Y. Hao, S. Hou, Q. Li, M. Liu, T. Mori, K. Prokofiev,
A. Salvucci, K. W. Tsang, H. Yang}

\AtlasCoverEdBoardMember{Lorenzo Bellagamba (chair)}
\AtlasCoverEdBoardMember{Pamela Ferrari}
\AtlasCoverEdBoardMember{Junjie Zhu}
\AtlasCoverEgroupEditors{atlas-EXOT-2015-20-editors@cern.ch}

\AtlasCoverEgroupEdBoard{atlas-EXOT-2015-20-editorial-board@cern.ch}

\hypersetup{pdftitle={ATLAS draft},pdfauthor={The ATLAS Collaboration}}

\begin{document}

\maketitle

\clearpage

\section{Introduction}
\label{sec:introduction}
Within the Standard Model (SM) of particle physics, direct production of lepton
pairs with different flavours ($\ell\ell^{\prime}$) is forbidden. 
However, lepton flavour violation (LFV) is allowed
in many extensions of the SM. Models with additional gauge symmetries, e.g.
production of a new heavy neutral gauge boson, similar to a \Zprime boson
\cite{bib:Langacker2008}, scalar neutrinos in $R$-parity-violating (RPV)
\cite{Farrar:1978xj,bib:RParitySUSY} supersymmetry (SUSY) \cite{Golfand:1971iw,Volkov:1973ix,Wess:1974tw,Wess:1974jb,Ferrara:1974pu,Salam:1974ig}, 
or low-scale gravity models predicting quantum black hole (QBH) production
\cite{bib:Gingrich2009} can produce decays to lepton-flavour-violating final
states. Processes leading to flavour-violating dilepton final states have a
clear detector signature and a low background from SM processes.
The Drell--Yan (DY) process (dilepton production in hadron--hadron collisions),
an irreducible background for same-flavour dilepton searches, is limited to the production
and decay of a ditau system, enhancing the sensitivity to a possible signal.
This paper looks for final states with two leptons of different flavour in
proton--proton ($pp$) collisions at $\sqrt{s}=13$ \si{\TeV}. The invariant
mass of the two leptons ($m_{\ell\ell^{\prime}}$) is used as the search variable.

A common extension of the SM is the addition of an extra $U(1)$ gauge symmetry
resulting in a massive vector boson known as a \Zprime boson
\cite{bib:Langacker2008}. The search presented in this paper assumes a \Zprime
boson that has the same fermion couplings as the SM \Z~boson in the quark
sector, but only leptonic decays that violate LFC are allowed. The addition of
lepton-flavour-violating processes, \Zprime$\to e\mu$, $e\tau$, $\mu\tau$,
requires new couplings between leptons of different generations: 
$Q^{\ell}_{\mathrm{12}}$, $Q^{\ell}_{\mathrm{13}}$ and
$Q^{\ell}_{\mathrm{23}}$, where the subscripts denote lepton generations. 
For the model considered, this paper assumes
$Q_{ij}^{\ell}$ equal to the SM \Z~boson coupling to one lepton and only one LFV
coupling different from zero at the same time. The ATLAS and CMS Collaborations
have placed limits on the $e\mu$, $e\tau$ and $\mu\tau$ couplings as a function
of the \Zprime boson mass up to
\SI{2.5}{\TeV}, using the full $\sqrt{s}=8$ \si{\TeV} \cite{bib:LFV8TeV,bib:CMS_8TeV_emu}.

In RPV SUSY, the Lagrangian terms allowing LFV can be expressed as
$\frac{1}{2}\lambda_{ijk}L_i L_j\bar{e_k} + \lambda^{\prime}_{ijk}L_i Q_j\bar{d_k}$,
where $L$ and $Q$ are the $SU(2)$ doublet superfields of leptons and quarks,
$e$ and $d$ are the $SU(2)$ singlet superfields of leptons and down-like quarks,
$\lambda$ and $\lambda^\prime$ are Yukawa couplings, and the indices $i$, $j$
and $k$ denote fermion generations. A $\tau$ sneutrino ($\tilde{\nu}_{\tau}$)
may be produced in $pp$ collisions by $d\bar{d}$ annihilation and subsequently
decay to $e\mu$, $e\tau$, or $\mu\tau$. Although only $\tilde{\nu}_{\tau}$ is
considered in this paper, results apply to any sneutrino flavour. For the theoretical
prediction of the cross-section times branching ratio, the $\tilde{\nu}_{\tau}$
coupling to first-generation quarks ($\lambda^{\prime}_{311}$) is assumed to be
0.11 for all channels. As for the \Zprime model, only one decay to a
lepton-flavour-violating final state is allowed at the same time. As such, for
an $e\mu$ final state, it is assumed that $\lambda_{312}$=$\lambda_{321}$=0.07,
for $e\tau$ $\lambda_{313}$=$\lambda_{331}$=0.07 and $\mu\tau$
$\lambda_{323}$=$\lambda_{332}$=0.07. These values are consistent with benchmark
couplings used in previous ATLAS and CMS searches \cite{bib:LFV8TeV,bib:CMS_8TeV_emu}. The ATLAS Collaboration has placed limits up to \SI{2.0}{\TeV} on the mass of an RPV SUSY $\tilde{\nu}_{\tau}$
\cite{bib:LFV8TeV}.

Various models introduce extra dimensions in order to lower the value of the
Planck mass ($M_{\mathrm{P}}$) and solve the hierarchy problem. The search
presented in this paper focuses on the ADD model~\cite{bib:Hamed1998}, assuming
$n=6$, where $n$ is the number of extra dimensions, and the RS model
\cite{bib:Randall1999}, with one extra dimension. Due to the increased strength
of gravity at short distances, $pp$ collisions at the Large Hadron Collider
(LHC) could produce states with masses beyond the threshold mass
($M_{\mathrm{th}}$), satisfying the Hoop conjecture~\cite{bib:Thorne1972} and
form black holes. For the model considered, $M_{\mathrm{th}}$ is assumed to be
equivalent to the extra-dimensional Planck scale. It is expected that, for
masses beyond 3-5$M_{\mathrm{th}}$, thermal black holes would be produced
\cite{bib:Dimopoulos2001,bib:Giddings2001}, characterised by high-multiplicity
final states. As such, for the search presented in this paper, it is more
interesting to focus on the mass region below 3--5$M_{\mathrm{th}}$, known as the
quantum gravity regime, investigated in Refs.
\cite{bib:MeadeRandall2007,bib:Gingrich2008,bib:Calmet2008}. Non-thermal (or
quantum) black holes would be formed in this region, and could decay to
two-particle final states, producing the topology this analysis is focused on.
Such quantum black holes would form a continuum in mass from $M_{\mathrm{th}}$ up
to the beginning of the thermal regime. For the model considered in this paper,
the thermal regime is assumed to start at 3$M_{\mathrm{th}}$. The decay of
quantum black holes would be governed by a yet unknown theory of quantum
gravity. The two main assumptions of the extra-dimensions models considered
\cite{bib:Gingrich2009} in this paper are:
\begin{itemize}
\item gravity couples with equal strength to all SM particle degrees of
  freedom;      
\item gravity conserves local symmetries (colour, electric charge) but can
  violate global symmetries such as LFC and baryon number conservation.
\end{itemize}
Following these assumptions, the branching ratio (BR) to each final state
can be calculated. Two initial states could give rise to a quantum black hole
decaying into a lepton-flavour-violating final state: $q\bar{q}$ and $gg$. The
branching ratio to $\ell\ell^{^{\prime}}$ is 0.87\% (0.34\%) for a $q\bar{q}$
($gg$) initial state \cite{bib:Gingrich2009}. This model was used in previous ATLAS and CMS searches in dijet
\cite{bib:DijetQBHPaper,bib:DijetQBHPaper13TeV, bib:CMS_8TeV_dijet}, lepton+jet
\cite{bib:LeptonJetQBHPaper}, photon+jet \cite{bib:PhotonJetQBHPaper}, $e\mu$
\cite{bib:CMS_8TeV_emu} and same-flavour dilepton \cite{bib:Matrix} final states.

\section{The ATLAS detector}
\label{sec:ATLASDetector}
The ATLAS detector \cite{bib:JINSTATLAS} is a general-purpose particle
detector with approximately forward-backward symmetric cylindrical
geometry.\footnote{ATLAS uses a right-handed coordinate system with its origin
  at the nominal interaction point (IP) in the centre of the detector and the
  $z$-axis along the beam pipe. The $x$-axis points from the IP to the centre
  of the LHC ring, and the $y$-axis points upward. The $x$--$y$ plane is
  referred to as the transverse plane, used to define quantities such as the
  transverse momentum ($p_{\mathrm{T}}$). Cylindrical coordinates $(r,\phi)$ are
  used in the transverse plane, $\phi$ being the azimuthal angle around the
  $z$-axis. The pseudorapidity is defined in terms of the polar angle $\theta$
  as $\eta=-\ln\tan(\theta/2)$. Angular distance is measured in units of
  $\Delta R = \sqrt{ (\Delta\eta)^{2} + (\Delta\phi)^{2} }$.} It is composed
of four main components, each responsible for identifying and reconstructing
different types of particles: the inner detector (ID), the electromagnetic
and hadronic calorimeters, and the muon spectrometer (MS).
Each of the sub-detectors is divided into two components, barrel and endcap, 
to provide coverage close to $4\pi$ in solid angle. In addition, two magnet
systems are in place to allow charge and momentum measurements: an axial
magnetic field of \SI{2.0}{\tesla} provided by a solenoid surrounding the ID,
and a toroidal magnetic field  for the MS.
The ID, the component of the ATLAS detector closest to the interaction point, 
reconstructs the trajectories of charged particles in the region $|\eta|<2.5$
and measures their momenta. It is composed of three sub-systems:
\begin{itemize}
\item[i)] a silicon pixel detector, including the newly installed insertable
B-layer \cite{bib:IBL_TDR,ATLAS:1451888};
\item[ii)] the semi-conductor tracker, used in conjunction with the silicon pixel detector
 to determine primary and secondary
vertices with high precision thanks to their high granularity;
\item[iii)] the transition radiation tracker, providing additional tracking in
the region $|\eta|<2.0$ and electron identification.
\end{itemize}
Surrounding the ID, lead/liquid-argon (LAr) sampling calorimeters provide
electromagnetic (EM) energy measurements with high granularity. A
steel/scintillator-tile hadronic calorimeter covers the central pseudorapidity
range ($|\eta| < $ 1.7). The endcap and forward regions are LAr calorimeters with 
copper or tungsten absorbers for both the EM and hadronic energy measurements
up to $|\eta| < $ 4.9. Built around the calorimeter system, the MS is the
sub-detector furthest from the interaction point. It consists of three layers
of precision tracking chambers and fast detectors for triggering on muons.
Tracking coverage is provided up to $|\eta| < $ 2.7 through the use of monitored
drift tubes  and, in the innermost layer, cathode strip chambers for
$|\eta| > $ 2.0, while trigger coverage is provided by resistive plate and
thin gap chambers up to $|\eta| < $ 2.4.

The trigger and data-acquisition system is based on two levels of online event
selection \cite{bib:TwoLevelTrigger}: the level-1 trigger and the high-level
trigger. The level-1 trigger is hardware-based and uses a subset of detector
information to provide quick trigger decisions and reduce the accepted rate to
\SI{100}{\kilo\hertz}. The high-level trigger is software-based and exploits
the full detector information to further reduce the accepted rate to about one
\si{\kilo\hertz}.

\section{Data and Monte Carlo simulated samples}
\label{sec:dataMC}
The data sample used for this analysis was collected with the ATLAS detector
during the 2015 LHC run with $pp$ collisions at a centre-of-mass energy of
\SI{13}{\TeV} with a \SI{25}{\nano\second} minimum proton bunch spacing. After
selecting periods with stable beams and requiring all detector systems to be
fully functional, the total integrated luminosity for the analysis is
\SI{3.2}{\per\femto\barn}. The uncertainty in the integrated luminosity is
5.0\%. It is derived following a methodology similar to that detailed in
Ref.~\cite{bib:LumiUncertainty}, from a calibration of the luminosity scale
using $x$--$y$ beam-separation scans performed in August 2015.

The $pp\to\Zprime\to\ell\ell^{\prime}$ signal samples are generated at leading
order (LO) using the Monte Carlo (MC) generator \Pythia~8.186~\cite{bib:Pythia8}
with the NNPDF23LO \cite{bib:Ball2012}  parton distribution function (PDF) set
and the A14~\cite{bib:A14tune} set of tuned parameters (tune). Signal samples 
with 25 mass points ranging
from \SI{0.5}{\TeV} up to \SI{5}{\TeV} are generated in \SI{0.1}{\TeV} steps
from 0.5 to \SI{2.0}{\TeV}, \SI{0.2}{\TeV} steps from 2.0 to \SI{3.0}{\TeV} and
\SI{0.5}{\TeV} steps from 3.0 to \SI{5.0}{\TeV}. The production cross-section is calculated
with the same MC generator used for simulation. No mixing with the SM \Z~boson
is included.

The $d\bar{d}\to\tilde{\nu}_{\tau}\to\ell\ell^{\prime}$ signal samples are
generated at LO using the MC generator \Madgraph~v2.3.3 \cite{bib:MadGraph5}
interfaced to the \Pythia~8.186 parton shower model with the NNPDF23LO PDF set
and the A14 tune. The signal samples are generated at the same
pole-masses as for the \Zprime described above. The cross-section is calculated 
at LO with the same MC generator used for simulation. A next-to-leading order (NLO)
correction factor ($K$-factor) is calculated for the cross-section based on  
Ref.~\cite{bib:RPVSUSYNLO} using \LoopTools~v2.2~\cite{bib:LoopTools}.

The $pp\to$QBH$\to\ell\ell^{\prime}$ samples are generated with \QBH~3.00
\cite{bib:QBHManual}  using the CTEQ6L1 \cite{bib:CTEQ6L1} PDF set and the A14 tune, for which
\Pythia~8.183 provides showering and hadronisation. For each extra-dimensional
model, eleven $M_{\mathrm{th}}$ points in 0.5~\TeV~steps were produced: from 3.0
to \SI{8.0}{\TeV} for the ADD $n=6$ model, and from 1.0 to \SI{6.0}{\TeV} for
the RS $n=1$ model. The production cross-section is calculated
with the same MC generator used for simulation. These two models have 
differences in the number and nature of the additional extra dimensions 
(large extra dimensions for ADD, one highly
warped extra dimension for RS). In particular, the ADD model allows production
of black holes with a larger gravitational radius and hence the parton--parton
cross-section for this model is larger than for the RS model. Therefore, the
$M_{\mathrm{th}}$ range of the generated samples is different for the two models.

The SM background to the LFV dilepton search is composed of several processes
which can produce a final state with two different-flavour leptons. The
dominant background contributions originate from \ttbar and single-top
production, with the subsequent decays of the top quark producing leptonically
decaying \W~bosons. Other backgrounds originate from diboson (\WW, \WZ~and \ZZ)
production and the DY process, $q\bar{q}\to \Z/\gamma^{*}\to\tau\tau$, which
can produce different-flavour final states through the leptonic decay of the
\W~and \Z~bosons and the $\tau$ lepton. Multi-jet and \W+jets processes
contribute due to the misidentification of jets as leptons.

Backgrounds from top quark production include \ttbar and single-top with an
associated \W~boson ($t\W$). Both the \ttbar and single-top-quark backgrounds
are generated at NLO using the \powhegbox~v2 \cite{bib:Alioli2010xd}
generator with the CT10 \cite{bib:Lai2010} PDF set in the matrix element (ME)
calculations. \Pythia~6.4.28 \cite{bib:Sjostrand2006za} and the corresponding
Perugia 2012 tune \cite{bib:Skands2010ak} are used to
simulate the parton shower, hadronisation, and the underlying event. Top quarks
are decayed using \madspin~\cite{bib:Artoisenet2012st}, preserving all spin
correlations. The parameter which controls the $p_{\mathrm{T}}$ of the first
emission beyond the Born configuration in Powheg, called \textit{hdamp}, is
set to the mass of the top quark. The main effect of this is to regulate the
high-$p_{\mathrm{T}}$ emission against which the \ttbar system recoils. The mass
of the top quark is set to \SI{172.5}{\GeV}. A value of 831$^{+20}_{-29}$ (scale)
$^{+35}_{-35}$ (PDF+$\alpha_{\mathrm{S}}$) $^{+23}_{-22}$ (mass uncertainty)
\si{\pico\barn} is used for the \ttbar production cross-section, computed with
\toppp2 \cite{bib:Czakon2011}, incorporating next-to-next-to-leading order
(NNLO) corrections in QCD, including resummation of next-to-next-to-leading
logarithmic (NNLL) soft gluon terms. A $t\W$ production cross-section of
$71.7\pm3.8$ \si{\pico\barn} is used, as computed in
Ref.~\cite{bib:SingleTopNNLO} to approximately NNLO (NNLL+NLO) accuracy.

Diboson processes with four charged leptons, three charged leptons and one
neutrino, two charged leptons and two neutrinos, or one boson decaying to
leptons and the other hadronically, are simulated using the~\Sherpa~2.1.1
generator \cite{bib:Gleisberg2008ta}. The matrix elements contain all diagrams
with four electroweak vertices. Fully-leptonic decays are calculated for up to
one (four leptons, two leptons and two neutrinos) or zero partons (three
leptons and one neutrino) at NLO and up to three partons at LO using the
\comix~\cite{bib:Gleisberg2008fv} and \openloops~\cite{bib:Cascioli2011va} ME
generators and merged with the \Sherpa~parton-shower \cite{bib:Schumann2007m}
using the ME+PS@NLO prescription \cite{Hoeche:2012yf}. Semileptonic decays are
calculated for up to one (\ZZ) or zero (\WW, \WZ) additional partons at NLO and
up to three additional partons at LO using~\comix~and~\openloops. The CT10 PDF
set is used in conjunction with the default parton-shower tuning provided by
the \Sherpa~authors in the release.

The Drell--Yan process is generated at LO using the \pythia8 MC generator
with the NNPDF23LO PDF set. The same generator is used for showering and
hadronisation. Dilepton mass-dependent $K$-factors are applied to account for
higher-order QCD and electroweak corrections and to normalise the cross-section
to NNLO, computed using \fewz~3.1 \cite{bib:FEWZ2012} and the CT14NNLO PDF set
\cite{bib:Dulat2015}.

SM processes such as \W+jets and multi-jet production involving jets that fake
leptons are evaluated through the use of data-driven methods detailed in
Section~\ref{sec:background}. The \W+jets contribution is estimated with the
aid of \Sherpa~MC simulated samples. Matrix elements are calculated for 
up to two partons at NLO and four partons at LO using
the same procedures, prescriptions and PDF set adopted for the diboson samples.
The \W+jets events are normalised to the NNLO cross-section \cite{bib:NNLO}.

For all samples used in this analysis, the effects of multiple interactions per
bunch crossing (pile-up) are accounted for by overlaying minimum-bias events
simulated with \pythia8 and re-weighting the MC events to reproduce the
distribution of the average number of interactions per bunch crossing observed
in the data. The MC generated events were processed with the ATLAS
simulation infrastructure~\cite{bib:ATLASSimul2010}, based on
\geant4~\cite{bib:Geant4}, and passed through the trigger simulation and the
same reconstruction software used for the data.

\section{Object and event selection}
\label{sec:obj_evt_selection}
Candidate muon tracks are initially reconstructed independently in the ID and
the MS. The two tracks are then used as input to a combined fit which takes
into account the energy loss in the calorimeter and multiple scattering.
Muon identification is based on information from both the ID and MS to ensure
that muons are reconstructed with the optimal momentum resolution up to very
high $p_{\mathrm{T}}$ using the \emph{High}-$p_{\mathrm{T}}$ operating point
\cite{Aad:2016jkr}. Muon candidates with hits in regions of the MS with
residual misalignments, such as the barrel--endcap overlap region
($1.01<|\eta|<1.1$), are vetoed. Muon tracks are required to be within the ID
acceptance region\footnote{For the $\mu\tau$ channel, the muon acceptance is
limited by the coverage of the muon trigger system ($|\eta|<2.4$).} of
$|\eta|<2.5$ and have at least three hits in each of the three traversed
precision chambers in the MS. An exception is made in the region $|\eta|<0.1$
due to the MS gap in that region, where tracks with at least three hits in a
single precision chamber are allowed. In order to suppress hadrons
misidentified as muons, the momentum measurements of the ID and the MS must
agree within seven standard deviations. As well as the quality cuts, muon
candidates must fulfil $p_{\mathrm{T}}>65$~\si{\GeV} and transverse impact
parameter ($d_{\mathrm{0}}$) significance $|d_{\mathrm{0}}/\sigma_{d_{\mathrm{0}}}|<3$
with respect to the beam line, where $\sigma_{d_{\mathrm{0}}}$ is the uncertainty in
the value of the transverse impact parameter. The distance between the
$z$-position of the point of closest approach of the muon track in the ID to
the beamline and the $z$-coordinate of the primary vertex\footnote{The primary
vertex corresponds to the interaction vertex with the highest $p_{\mathrm{T}}^2$
sum of all tracks belonging to it.} ($\Delta z_{\mathrm{0}}$) is required to
satisfy $|\Delta z_{\mathrm{0}}\sin\theta|<0.5$~\si{\milli\meter}. This
requirement aims to reduce the background from cosmic rays and from muons
originating from heavy-flavour decays. Moreover, candidates are required to
fulfil track-based isolation criteria with a fixed efficiency of 99\% over the
full range of muon momentum to further reduce contamination from  non-prompt
muons. The sum of the transverse momentum of tracks in an isolation cone of
size $\Delta R=0.2$ (excluding the muon itself) divided by the muon
$p_{\mathrm{T}}$ is used as a discrimination criterion for the track-based
isolation.

Electron candidates are formed from the energy in clusters of cells in the
electromagnetic calorimeter associated with a track in the ID
\cite{bib:electrons2014}. A multivariate analysis approach, employing a
likelihood (LH) discriminant, is built to suppress contributions from hadronic
jets, photon conversions, Dalitz decays and semileptonic heavy-flavour hadron
or kaon decays. The LH discriminant utilises lateral and longitudinal
calorimeter shower shape, tracking and cluster--track matching quantities. The
discriminant criterion is a function of the tranverse momentum and $|\eta|$ of
the candidate electron. Two operating points are used in this analysis, as
defined in Ref.~\cite{bib:LHIdentification}: \emph{Medium} and \emph{Tight}.
The \emph{Tight} working point (90\% efficient at $p_{\mathrm{T}}=65$~\si{\GeV})
is required for electron candidates, while the \emph{Medium} working point
(95\% efficient at $p_{\mathrm{T}}=65$~\si{\GeV}) is used to estimate the
background contribution from jets misidentified as electrons (as discussed in
Section \ref{sec:background}). Electron candidates must fulfil
$p_{\mathrm{T}}>65$~\si{\GeV} and $|\eta|<2.47$, excluding the region
$1.37<|\eta|<1.52$, where the energy reconstruction performance is degraded due
to the presence of extra inactive material. Further requirements are made on
the impact parameter: $|d_{\mathrm{0}}/\sigma_{d_{\mathrm{0}}}|<5$ and
$|\Delta z_{\mathrm{0}}\sin\theta|<0.5$~\si{\milli\meter}. To reject electrons
faked by muons, electron candidates within a $\Delta R=0.2$ cone around a muon
candidate are removed. Moreover, candidates are required to fulfil relative
track- (as defined above for muon candidates) and calorimeter-based isolation
requirements with a fixed efficiency of 99\%, to suppress background from
non-prompt leptons originating from heavy-flavour or kaon decays, charged
hadrons and photon conversions from $\pi^{\mathrm{0}}$ decays. The sum of the calorimeter
transverse energy deposits in an isolation cone of size $\Delta R=0.2$
(excluding the electron itself) divided by the electron $p_{\mathrm{T}}$ is used
as a discrimination criterion for the calorimeter-based isolation.

Jets, used in the reconstruction of hadronically-decaying $\tau$ leptons, are reconstructed using the anti-$k_{t}$ algorithm \cite{bib:Cacciari2008}
with a radius parameter ($R$) of 0.4, using as input topological clusters
\cite{bib:TopoClusters} of calorimeter cells \cite{bib:JetCalibrations}. The
three-dimensional topological clusters are built from topologically connected
calorimeter cells that contain a significant signal above noise. The cluster
energies are corrected for inactive material and out-of-cluster energy losses.
Jet calibrations derived from $\sqrt{s}=13$~\si{\TeV} simulation, and collision
data taken at $\sqrt{s}=8$~\si{\TeV} and $\sqrt{s}=13$~\si{\TeV}, are used to
correct the jet energies and directions to those of the particles from the
hard-scatter interaction. This calibration procedure, described in
Refs.\cite{bib:JetCalibrations,bib:PileupJets,bib:JetCorrections8TeV}, is
improved by a data-derived correction to the relative calibration of jets in
the central and the forward regions.

The reconstruction of $\tau$ leptons and their visible hadronic decay products,
referred to as $\tau_{\mathrm{had}}^{\mathrm{vis}}$,  starts with jets reconstructed
from topological clusters as described above. Hadronic
decays of $\tau$ leptons ($\tau_{\mathrm{had}}$) are mainly characterised
by the presence of one or three charged particles, accompanied by a neutrino
and possibly other neutral particles \cite{bib:TauReconstructionRun2}. The
$\tau_{\mathrm{had}}^{\mathrm{vis}}$ candidates must have energy deposits in the
calorimeters in the range $|\eta|<2.5$, with the transition region between the
barrel and endcap calorimeters ($1.37<|\eta|<1.52$) excluded, a transverse
momentum greater than \SI{40}{\GeV}, one or three associated tracks and an
electric charge of $\pm$1. Their identification is performed using a
multivariate algorithm that employs boosted decision trees (BDTs) to
discriminate against quark- and gluon-initiated jets using shower shape and
tracking information. An additional dedicated likelihood-based veto is used to
reduce the number of electrons misidentified as $\tau_{\mathrm{had}}$. The $\tau$
lepton candidates which overlap with electron or muon candidates within a cone
of $\Delta R=0.2$ are rejected.

The event selection requires a single-muon or single-electron trigger with a
$p_{\mathrm{T}}$ threshold of \SI{50}{\GeV} for muons, and 60 or \SI{120}{\GeV}
for electrons. The single-electron trigger with higher $p_{\mathrm{T}}$ threshold
has a looser LH identification requirement, resulting in an increased trigger
efficiency at high $p_{\mathrm{T}}$. Selected events must have a reconstructed
primary vertex and exactly two different-flavour lepton candidates meeting the
above-mentioned criteria. Events with an additional lepton or extra ``loose''
lepton\footnote{A loose lepton is defined as a lepton satisfying all
requirements except isolation for muons and a looser identification requirement
(\textit{LH-Medium}) for electrons. No loose $\tau$ lepton category is
defined.} are vetoed. Moreover, the lepton candidates have to be back-to-back
in the $\phi$ direction with $\Delta\phi(\ell,\ell^{\prime})>2.7$. No requirement
is made on the respective charges of the leptons as it is found to reduce the
signal efficiency by as much as 6\% for the highest-mass signals considered due
to charge mis-assignment, without a significant effect on the background
rejection. For a \Zprime boson with a mass of \SI{1.5}{\TeV}, the acceptance
times efficiency\footnote{The acceptance ($A$) defines the geometrical and kinematic
region covered by the detector. The efficiency ($\epsilon$) is the fraction of events
falling in the detector acceptance region that fulfil all selection criteria.
Therefore, $A\epsilon$ is the fraction of events that pass all the selection
requirements.} ($A\epsilon$) of the selection requirements is approximately
50\%, 25\% and 20\% for the $e\mu$, $e\tau$ and $\mu\tau$ final states,
respectively. To account for differences between data and simulation,
corrections are applied to the lepton trigger, reconstruction, identification,
and isolation efficiencies as well as the lepton energy/momentum resolution
and scale \cite{Aad:2016jkr,bib:electrons2014,bib:TauReconstructionRun2}.

The missing transverse momentum ($E_{\mathrm{T}}^{\mathrm{miss}}$) is defined as the
negative vector sum of the transverse momenta of all identified physics objects
(electrons, photons \cite{bib:PhotonReconstruction}, muons, taus, jets) and an additional soft term. The soft term is constructed from all tracks that are associated with the primary vertex
but not with any physics object. In this way, the missing transverse momentum
is adjusted for the best calibration of the jets and the other identified
physics objects above, while maintaining pile-up independence in the soft term
\cite{bib:METPerformance13TeV}.

An additional variable to estimate the contribution from reducible backgrounds
is used: the transverse mass
($m_{\mathrm{T}}$) of a lepton and the $E_{\mathrm{T}}^{\mathrm{miss}}$, defined as:
\begin{equation}
  \label{eq:mT}
  m_{\mathrm{T}} = \sqrt{2 p_{\mathrm{T}}E_{\mathrm{T}}^{\mathrm{miss}}
                 (1-\cos(\Delta\phi(\ell,E_{\mathrm{T}}^{\mathrm{miss}}))}\;,
\end{equation}
where $\Delta\phi(\ell,E_{\mathrm{T}}^{\mathrm{miss}})$ is the azimuthal angle
between the lepton $p_{\mathrm{T}}$ and $E_{\mathrm{T}}^{\mathrm{miss}}$ direction.

For events in the $e\tau$ and $\mu\tau$ channels, in order to reconstruct the
dilepton invariant mass more accurately, the neutrino four-momentum is taken
into account. The hadronic decay of a $\tau$ lepton from a heavy resonance
leads to the neutrino and the resultant jet being nearly collinear. The
neutrino four-momentum is reconstructed from the magnitude of the missing
transverse momentum, and is assumed to be collinear with the $\tau_{\mathrm{had}}$
candidate. For the mentioned channels, the above technique significantly
improves the mass resolution and search sensitivity.

\section{Background estimation}
\label{sec:background}
The background processes for this search can be divided into two categories:
irreducible and reducible backgrounds. The former is composed of processes
which can produce two different flavour prompt leptons in the final state,
including the DY$\to\tau\tau$ process, \ttbar, single top, and diboson
production. These processes are modelled using MC simulated samples. Reducible
backgrounds occur when jets are mis-reconstructed as leptons, and require the
use of data-driven techniques.

The MC samples used to estimate single-top and \ttbar production are
statistically limited for dilepton invariant masses above 1~\TeV. Therefore,
fits to the $m_{\ell\ell^{\prime}}$  distribution using monotonically decreasing
functions are used to extrapolate those backgrounds to the region
$m_{\ell\ell^{\prime}}>1$~\si{\TeV}. Two functional forms are investigated, chosen
for their stability when varying the fit range and for the quality of the fit:
\begin{equation}
\label{eq:djet-monimial}
\mathrm{e}^{-a} \cdot m_{\ell\ell^{\prime}}^b \cdot
m_{\ell\ell^{\prime}}^{c\cdot\mathrm{ln}(m_{\ell\ell^{\prime}})}
\quad \mathrm{and} \quad \frac{a}{(m_{\ell\ell^{\prime}}+b)^c}\;,
\end{equation}
where $a$, $b$ and $c$ are free parameters in the fit. A study of the
stability of the fit was performed by varying the lower and upper limits of
the fit range between 200--\SI{300}{\GeV} and 1000--1200~\GeV~in
\SI{25}{\GeV} steps, respectively. The stitching point between the MC
estimation and the fit is chosen to be at \SI{900}{\GeV} for the top quark
background. The nominal extrapolation is then taken to be the median of all the
tested fit ranges using both functional forms. Good agreement is found between
the fit prediction and the available MC events. The addition in quadrature of
the fit parameter uncertainties and the RMS of all fit variations is assigned
as a systematic uncertainty. 

The contribution from reducible backgrounds originate mainly from \W+jets and
multi-jet processes. The background of muons originating from hadronic decays
is found to be negligible compared to the contribution from fake electrons and
taus. Therefore, in the $e\mu$ channel, where the contribution of the reducible
background is expected to be small, these non-prompt muons are neglected. The
reducible background in that channel is then reduced to events with one prompt
muon and a jet faking an electron. This background contribution is usually not
well modelled by MC simulation.

For the $e\mu$ channel, a technique known as the matrix method, described in
Ref.~\cite{bib:Matrix}, is employed. Exclusive samples are defined by
loosening the selection criteria for electron candidates. Here the matrix
method involves two parameters that need to be determined as a function of
electron $p_{\mathrm{T}}$: the probability of a loose electron to pass the full
object selection, the so-called \emph{real electron efficiency}
($\epsilon_{\mathrm{R}}$), and the probability of a jet fulfilling the loose
electron selection criteria to pass the full selection, known as the
\emph{electron fake rate} ($\epsilon_{\mathrm{F}}$). The former is evaluated from
MC simulation, while the latter is evaluated in a data sample dominated by
multi-jet events. To construct this multi-jet control sample, it is required
that $E_{\mathrm{T}}^{\mathrm{miss}}<25$~\si{\GeV} and $m_{\mathrm{T}}<50$~\si{\GeV}
in order to suppress the \W+jets contribution. Contamination from \W+jets and
other SM background processes (top, diboson, and $Z\to\ell\ell$) is subtracted
using MC predictions. 

For the $e\tau$ and $\mu\tau$ channels, the $\tau$ fake rate is measured in
data in a \W$\to e$/$\mu$+jets control region as a function of the
$\tau_{\mathrm{had}}^{\mathrm{vis}}$ $p_{\mathrm{T}}$. The region is defined to be
orthogonal to the signal selection by reversing the
$\Delta\phi(\ell,\ell^{\prime})$ requirement. Only events with exactly one
electron or muon fulfilling all selection criteria (as defined in 
Section~\ref{sec:obj_evt_selection}), as well as $m_{\mathrm{T}}>60$~\si{\GeV},
are used. The $\tau_{\mathrm{had}}$ candidates present in those events are
dominated by jets. The $\tau$ fake rate is defined as the fraction of jets
fulfilling all $\tau$ object selection criteria, including the multivariate
BDT-based identification. The derived fake rate is used to weight simulated
\W+jets events. After obtaining the fake-rate-weighted $m_{\ell\ell^{\prime}}$
distribution, a normalisation factor for the \W+jets background is obtained in
a \W+jets enriched region to scale the overall normalisation of the MC simulation
to that of the data. The \W+jets enriched region is defined as a sub-set of the signal selection by
further requiring $E_{\mathrm{T}}^{\mathrm{miss}}>30$~\si{\GeV} and lepton
$p_{\mathrm{T}}<150$~\GeV~to avoid possible signal contamination. The
contribution from events with an electron/muon and a fake $\tau_{\mathrm{had}}$
is found to make up around 55\% of the overall background in the $e\tau$ and
$\mu\tau$ channels.

To evaluate the fake background from events with a real $\tau_{\mathrm{had}}$ and
a fake electron/muon in the $e\tau$ and $\mu\tau$ channels, a
fake-electron/muon enriched sample is defined by requiring a non-isolated
electron/muon and a $\tau_{\mathrm{had}}$ candidate. Three regions are defined:
\begin{itemize}
\item[] \textbf{Region 1:} pairs of a non-isolated electron/muon and a
$\tau_{\mathrm{had}}$ with the same electric charge;
\item[] \textbf{Region 2:} pairs of an isolated electron/muon and a
$\tau_{\mathrm{had}}$ with the same electric charge;
\item[] \textbf{Region 3:} all pairs of a non-isolated electron/muon and a
$\tau_{\mathrm{had}}$.
\end{itemize}
The $m_{\ell\ell^{\prime}}$ shape of the  contribution is obtained from region 3 by
subtracting the contribution from other background sources to the data, while
the ratio of isolated to non-isolated leptons in regions 1 and 2 is used to
normalise this background contribution appropriately. The contribution from
events with a fake electron/muon and a real $\tau$ lepton is found to be below
1\% in the $\mu\tau$ channel, while in the $e\tau$ channel its contribution to
the overall SM background is close to 5\%.

A summary of the contribution from each SM background in each of the final
states can be found in Section~\ref{sec:results}.

\section{Systematic uncertainties}
\label{sec:systematics}
Sources of systematic uncertainty are divided in two categories: theoretical
and experimental. Uncertainties in the predicted cross-section times branching
ratio and the modelling of the $m_{\ell\ell^{\prime}}$ shape  
of the background processes considered are regarded  as theoretical
uncertainties, while uncertainties relating to the simulation of the detector
response are regarded as experimental uncertainties. Theoretical uncertainties
(such as PDF-related uncertainties) in the signal cross-section are not
considered in this paper.

The PDF uncertainties are the dominant theoretical systematic uncertainties,
together with the uncertainty of the extrapolation to estimate the background
contribution at high-mass (as described in Section \ref{sec:background}). The
contribution from PDF uncertainties is estimated using different PDF sets and
eigenvector uncertainty sets within a particular PDF. The CT10 PDF
uncertainty due to eigenvector variations is evaluated
through the use of LHAPDF \cite{bib:LHAPDF2014} following the prescriptions
outlined in Ref.~\cite{bib:PDF4LHC}. The uncertainty related to the choice of
PDF is evaluated by comparing the results with those from the central value of
other PDF sets such as MMHT2014 \cite{bib:Motylinski2014sya}, NNPDF3.0
\cite{bib:Ball2014uwa} and CT14 \cite{bib:Dulat2015}. PDF-related uncertainties
in the signal shape are not considered. The uncertainties in the the
$m_{\ell\ell^{\prime}}$ modelling in \ttbar events is obtained using separate MC
samples generated with variations in the renormalisation and
factorisation scales and the \textit{hdamp} parameter (as defined in 
Section~\ref{sec:dataMC}).

The effect of experimental systematic uncertainties is assessed through the
uncertainties associated to the corrections applied to simulated processes,
including lepton momentum resolution and scale, and trigger, identification,
reconstruction and isolation efficiencies
\cite{Aad:2016jkr,bib:electrons2014,bib:TauReconstructionRun2}. The
efficiencies are evaluated using events from the $\Z\to\ell\ell$ peak and then
extrapolated to high energies.

Mismodelling of the muon momentum resolution at the \si{\TeV} scale, such as due to
residual misalignment of the muon precision chambers, can alter the signal and
background shapes. An uncertainty related to this is obtained from studies
performed in dedicated data-taking periods with no magnetic field in the MS.
The muon reconstruction efficiency is affected at high-$p_{\mathrm{T}}$ by
possible large energy losses in the calorimeter. The associated uncertainty is
estimated by comparing studies with $\Z\to\mu\mu$ events in data extrapolated
at high-$p_{\mathrm{T}}$ to the results predicted by MC simulation
\cite{bib:Dilepton13TeV}. The effect on the muon reconstruction efficiency was
found to be approximately 3\% per \si{\TeV} as a function of muon
$p_{\mathrm{T}}$.

The uncertainty in the electron identification efficiency extrapolation is
based on the differences in the electron shower shapes in the EM calorimeters
between data and MC simulation in the $\Z\to ee$ peak, which are propagated to
the high-$p_{\mathrm{T}}$ electron sample. The effect on the electron
identification efficiency was found to be 2\% and is independent of
$p_{\mathrm{T}}$ for electrons with transverse momentum above \SI{150}{\GeV}
\cite{bib:Dilepton13TeV}.

The treatment of systematic uncertainties for $\tau$ leptons with
$p_{\mathrm{T}}$ up to \SI{100}{\GeV} is detailed in
Ref.~\cite{bib:TauReconstructionRun2}. An additional uncertainty of 
20\% per TeV is assigned to the reconstruction efficiency of $\tau$ leptons 
with pT $>$ \SI{100}{\GeV} to account for the 
the degradation of the modelling and reconstruction efficiency due to track
merging, derived through studies in simulation and in dijet data events at \SI{8}{\TeV}
\cite{bib:TauReconstructionRun1}. 

The uncertainties associated to the matrix method used for the $e\mu$ channel 
are evaluated by considering effects on the $\epsilon_{\mathrm{F}}$ measurement,
including the multi-jet control sample definition and the uncertainties in
the overall normalisation. The former effect is evaluated by shifting the
$E_{\mathrm{T}}^{\mathrm{miss}}$ and $m_{\mathrm{T}}$ requirements by
$\pm$\SI{10}{\GeV}, while the latter is taken into account by varying the MC
subtraction of other SM processes by the luminosity and experimental systematic
uncertainties. For the $e\tau$ and $\mu\tau$ channels, the uncertainty in the
$\tau$ fake rate and \W+jets normalisation in the MC subtraction is considered.
The $\tau$ fake rate is re-evaluated when removing the $m_\mathrm{T}$
requirement, requiring $m_{\tau\ell}>110$~\si{\GeV} to reduce the Drell--Yan
background and vetoing events with a jet identified as originating from a
$b$-quark \cite{bib:btagging2015} to reduce top-quark background contamination.
The variations obtained for the $\tau$ fake rates are assigned as systematic
uncertainties. Given the limited data available for $\tau$ lepton
$p_{\mathrm{T}}>500$~\si{\GeV}, the statistical uncertainty from the last data
bin is used together with an uncertainty of 20\% per \si{\TeV} in $\tau$ lepton
$p_{\mathrm{T}}$. The uncertainty on the \W+jets normalisation is obtained by
recalculating the normalisation factor after a variation for each of the
experimental systematic uncertainties outlined in Table~\ref{tab:sig_syst}.

The uncertainty in the reducible background estimate is found to be close to
50\%, 30\% and 40\% for the $e\mu$, $e\tau$ and $\mu\tau$ channels,
respectively, at $m_{\ell\ell^{\prime}}=1.0$~\si{\TeV} and it is of comparable size
to the PDF uncertainty in the $e\tau$ and $\mu\tau$ channels. However, the
contribution from reducible backgrounds in the $e\mu$ channel is below 10\%,
while for $e\tau$ and $\mu\tau$ final states it is the leading background
together with the contribution from top quark production.

Experimental systematic uncertainties common to signal and background processes
are assumed to be correlated. The effect of
systematic uncertainties on the estimated SM background yields is summarised in
Table~\ref{tab:sig_syst}.

For signal processes, only experimental systematic uncertainties are
considered. The statistical uncertainty of the signal MC samples is 3\%.
\begin{table}[!htbp]
  \centering
  \footnotesize
   \begin{tabular}{l|rrr|rrr|rrr}
    \hline \hline       
    \multirow{2}{*}{Source}
    & \multicolumn{3}{c|}{$m_{\ell\ell^{\prime}}=1$~\si{\TeV}}
    & \multicolumn{3}{c|}{$m_{\ell\ell^{\prime}}=2$~\si{\TeV}}
    & \multicolumn{3}{c}{$m_{\ell\ell^{\prime}}=3$~\si{\TeV}}\\
    \cline{2-10}
    & $e\mu$ & $e\tau$ & $\mu\tau$ & $e\mu$ & $e\tau$ & $\mu\tau$
    & $e\mu$ & $e\tau$ & $\mu\tau$\\
    \hline
    PDF uncertainty                       & 17\% & 15\% & 15\% & 35\% & 38\% & 35\% & 70\%  & 75\%  & 70\%  \\
    Luminosity                            & 5\%  & 5\%  & 5\%  & 5\%  & 5\%  & 5\%  & 5\%   & 5\%   & 5\%   \\
    Statistical                           & 18\% & 11\% & 15\% & 80\% & 27\% & 27\% & 120\% & 28\%  & 30\%  \\
    Reducible background                  & 5\%  & 29\% & 40\% & 5\%  & 35\% & 75\% & 5\%   & 45\%  & 85\%  \\
    Top quark production modelling        & 5\%  & 3\%  & 4\%  & 12\% & 4\%  & 5\%  & 15\%  & 10\%  & 8\%   \\
    \hline
    Electron trigger  efficiency          & 1\%  & 1\%  & N/A  & 1\%  & 1\%  & N/A  & 1\%   & 1\%   & N/A   \\
    Electron identification               & 2\%  & 2\%  & N/A  & 2\%  & 2\%  & N/A  & 2\%   & 2\%   & N/A   \\
    Electron energy scale and resolution  & 3\%  & 3\%  & N/A  & 3\%  & 3\%  & N/A  & 3\%   & 3\%   & N/A   \\
    \hline
    Muon reconstruction efficiency        & 2\%  & N/A  & 2\%  & 4\%  & N/A  & 4\%  & 6\%   & N/A   & 6\%   \\
    Muon scale and resolution             & 4\%  & N/A  & 4\%  & 12\% & N/A  & 12\% & 20\%  & N/A   & 20\%  \\
    Muon trigger  efficiency              & 2\%  & N/A  & 2\%  & 2\%  & N/A  & 2\%  & 2\%   & N/A   & 2\%   \\
    \hline
    Tau identification                    & N/A  & 4\%  & 4\%  & N/A  & 5\%  & 5\%  & N/A   & 6\%   & 6\%   \\
    Tau reconstruction                    & N/A  & 3\%  & 3\%  & N/A  & 4\%  & 4\%  & N/A   & 4\%   & 4\%   \\
    Tau energy calibrations               & N/A  & 2\%  & 2\%  & N/A  & 3\%  & 3\%  & N/A   & 4\%   & 4\%   \\
    \hline
    Total                                 & 27\% & 35\% & 44\% & 90\% & 59\% & 90\% & 140\% & 90\%  & 120\% \\
    \hline \hline
    SM Background in $m_{\ell\ell^{\prime}}\pm0.1\cdot m_{\ell\ell^{\prime}}$
                                          & 3.9  & 11.9 & 11.4 & 0.09 & 0.55 & 0.49 & 0.002 & 0.014 & 0.017 \\
    \hline \hline
  \end{tabular}

  \caption{Quantitative summary of the systematic uncertainties taken into
    account for background processes. Values are provided for
    $m_{\ell\ell^{\prime}}$ values of 1, 2 and \SI{3}{\TeV}. The statistical error
    includes the extrapolation uncertainties of the top quark background in the
    high-$m_{\ell\ell^{\prime}}$ region together with the uncertainty related to the
    number of MC events. Uncertainties are quoted with respect to the total
    background. N/A means the systematic uncertainty is not applicable. The
    expected SM background in a mass window within
    $\pm0.1\cdot m_{\ell\ell^{\prime}}$ is also reported.}
  \label{tab:sig_syst}
\end{table}

\section{Statistical analysis}
\label{sec:statistic}
If no deviations from the SM prediction are observed, model-dependent
exclusion limits are extracted using a Bayesian method and implemented with
the software package Bayesian Analysis Toolkit (BAT) \cite{bib:Caldwell2008}
using a template shape method. A binned likelihood function ($\mathcal{L}$) is
built as the product of the Poisson probability of observing $n_{\mathrm{obs}_k}$
when expecting $\mu_{k}$ in each of the mass bins used
for the search:
\begin{equation}
  \label{eq:ProdLikelihood}
  \centering
    \mathcal{L}(n_{\mathrm{obs}}| \theta,\hat{\Omega}) =
    \prod^{N_{\mathrm{bins}}}_{k=1}
    \dfrac{\mu^{n_{\mathrm{obs}_k}}_{k}\mathrm{e}^{\mu_{k}}}{n_{\mathrm{obs}_{k}}!}
    \prod_{i=1}^{N_{\mathrm{Sys}}} G(\Omega_{i},0,1)\;,
\end{equation}
where $\mu_{k}$ is the expected number of background and signal events
($\mu_{k}=N_{\mathrm{bkg}_k} + N_{\mathrm{sig}_k}(\theta)$) as a function of the
parameter of interest $\theta$, $\hat{\Omega}$ is the vector of nuisance
parameters introduced to account for the effect of systematic uncertainties in
the expected yields, $N_{\mathrm{bins}}$ is the number of dilepton invariant mass
bins, $N_{\mathrm{Sys}}$ is the total number of nuisance parameters and
$G(\Omega_{i},0,1)$ is a Gaussian distribution with zero mean and unit standard
deviation assumed to be the probability density function for the nuisance
parameter $\Omega_{i}$. The dependence on the vector of nuisance parameters is
removed through the use of a Markov Chain Monte Carlo integration technique.
Bayes theorem is then applied to construct a posterior probability density
function for the number of signal events assuming a uniform prior in the
parameter of interest ($P(\theta)$). The number of signal events can be
expressed in terms of the cross-section times branching ratio of the signal process
($\sigma\cdot\mathrm{BR}(X\to \ell\ell^{\prime})$) as:
\begin{equation}
  \label{eq:NsigEvt}
  N_{\mathrm{sig}} = \sum^{\mathrm{N_{\mathrm{bins}}}}_{k=1} N_{\mathrm{sig}_{k}}
  = \sigma\cdot \mathrm{BR}(X\to \ell\ell^{\prime})\cdot L \cdot
  A\epsilon(X\to  \ell\ell^{\prime})\;,
\end{equation}
where $L$ is the integrated luminosity of the dataset and
$A\epsilon(X\to \ell\ell^{\prime})$ is the acceptance times efficiency of the
physics model tested. As such, a posterior probability density function is
obtained for the signal $\sigma\cdot\mathrm{BR}$. A 95\% credibility level (CL)
upper limit is obtained on the signal cross-section times branching ratio
 by finding the value of $\theta^{\mathrm{95}}$ satisfying:
\begin{equation}
  \label{eq:95CL}
  0.95 = \dfrac{\int^{\theta^{\mathrm{95}}}_{\mathrm{0}}
  \mathcal{L^{\prime}}(n_{\mathrm{obs}}|\theta)P(\theta)\mathrm{d}\theta }
  {\int^{\infty}_0\mathcal{L^{\prime}}(n_{\mathrm{obs}}|\theta)P(\theta)
  \mathrm{d}\theta}\;,
\end{equation}
where $P(\theta)$ is the uniform prior probability mentioned above and
$\mathcal{L^{\prime}}$ is the marginalised likelihood, obtained after performing
the Markov Chain Monte Carlo integration over $\hat{\Omega}$. Expected
exclusion limits are obtained by running 1000 pseudo-experiments (PE) for each
of the signal mass points tested. The median value of the 95\% CL upper
Bayesian limit PE distribution is taken as the expected limit. The one- and
two-standard deviation intervals of the expected limit are obtained from the
1000 PE ensemble by finding the 68\% and 95\% CL interval envelopes,
respectively.

The predicted width of the \Zprime boson, 3\% for $m_{\Zprime}=2$~\si{\TeV}, is
lower than the detector resolution for the $e\mu$ and the $\mu\tau$ channels,
which are approximately 8\% and 12\%, respectively, at the same \Zprime boson
mass. For the $e\tau$ final state the detector resolution is 4\% at
$m_{\Zprime}=2$~\si{\TeV}, comparable to the \Zprime boson width. The width of
the $\tilde{\nu}_{\tau}$ is below 1\% and hence the resolution of the detector
is larger than the width for each of the final states investigated. For limit setting on the
signal models investigated, a
logarithmic $m_{\ell\ell^{\prime}}$ binning is used 
with 40 mass bins between 120 and \SI{10000}{\GeV}. The bin width is
around 10\% in dilepton mass throughout the whole range.

\section{Results}
\label{sec:results}
Table~\ref{tab:yields} summarises the expected and observed yields in the
validation and search regions for each of the channels considered in this
search. The region $m_{\ell\ell^{\prime}}<600$~\si{\GeV} is defined as the
validation region where the data is used to check the SM background prediction,
while the region $m_{\ell\ell^{\prime}}>600$~\si{\GeV} is defined as the search
region. Selected $e\mu$ events are dominated by \ttbar events, while \W+jets
events are dominant for the $e\tau$ and $\mu\tau$ final states.
\begin{table}[!h]
  \centering
  \footnotesize
  \subfloat[$e\mu$ channel]{
      \begin{tabular}{l|r@{\,}@{$\pm$}@{\,}l@{\,}|r@{\,}@{$\pm$}@{\,}l@{\,}}
      \hline\hline
      Process               & \multicolumn{2}{c|}{$m_{e\mu}<600$~\si{\GeV}}
      &  \multicolumn{2}{c}{$m_{e\mu}>600$~\si{\GeV}}\\
      \hline
      Top quark             & \;\;\;1190 &  140
                            & \;\;\;\;\;\;22 & 5 \\
      Diboson               &  159 &   17
                            &  4.9 &  0.9  \\
      Multi-jet and \W+jets &   55 &   11
                            &  2.7 &  1.7  \\
      \DY                   & 14.5 &  2.0
                            & 0.18 & 0.04  \\
      \hline
      Total SM background   & 1410    & 150
                            &   30    &   7   \\
      \hline
      SM+\Zprime ($M_{\Zprime}=2$~\si{\TeV})
      & \multicolumn{2}{c|}{-} &  75 & 13     \\
      SM+$\tilde{\nu}_{\tau}$ ($M_{\tilde{\nu}_{\tau}}=2$~\si{\TeV})
      & \multicolumn{2}{c|}{-} &  40 &  8     \\
      SM+QBH RS $n = 1$ ($M_{\mathrm{th}}=2$~\si{\TeV})
      & \multicolumn{2}{c|}{-} &  44 &  9     \\
      \hline
      Data                  & \multicolumn{2}{c|}{1463}
                            & \multicolumn{2}{c}{25}    \\
      \hline\hline
    \end{tabular}
  }\\
  \subfloat[$e\tau$ channel]{
   \begin{tabular}{l|r@{\,}@{$\pm$}@{\,}l@{\,}|r@{\,}@{$\pm$}@{\,}l@{\,}}
      \hline\hline
      Process               & \multicolumn{2}{c|}{$m_{e\tau}<600$~\si{\GeV}}
      &  \multicolumn{2}{c}{$m_{e\tau}>600$~\si{\GeV}}\\
      \hline
      Top quark             &  \;\;\;\;\;790   &  190
                            &  \;\;\;\;\;\;25   &    9   \\
      Diboson               &  109   &   26
                            &    6.2 &    1.9 \\
      Multi-jet and \W+jets & 3200   &  800
                            &   45   &   14   \\
      \DY                   & 1030   &  240
                            &    5.2 &    1.4 \\
      \hline
      Total SM background   & 5200   & 1300
                            &   81   &   25   \\
      \hline
      SM+\Zprime ($M_{\Zprime}=1.5$~\si{\TeV})
      & \multicolumn{2}{c|}{-} & 185 &   34   \\
      SM+$\tilde{\nu}_{\tau}$ ($M_{\tilde{\nu}_{\tau}}=1.5$~\si{\TeV})
      & \multicolumn{2}{c|}{-} & 105 &   27   \\
      SM+QBH RS $n = 1$ ($M_{\mathrm{th}}=1.5$~\si{\TeV})
      & \multicolumn{2}{c|}{-} & 122 &   28   \\
      \hline
      Data                  &  \multicolumn{2}{c|}{5416}
                            &  \multicolumn{2}{c}{111}   \\
      \hline\hline
    \end{tabular}
  }\\
  \subfloat[$\mu\tau$ channel]{
    \begin{tabular}{l|r@{\,}@{$\pm$}@{\,}l@{\,}|r@{\,}@{$\pm$}@{\,}l@{\,}}
     \hline\hline
      Process               & \multicolumn{2}{c|}{$m_{\mu\tau}<600$~\si{\GeV}}
      &  \multicolumn{2}{c}{$m_{\mu\tau}>600$~\si{\GeV}}\\
      \hline
      Top  quark            &  \;\;\;\;\;580   & 140
                            &  \;\;\;\;\;\;21   &   7   \\
      Diboson               &   84   &  20
                            &    4.8 &   1.4 \\
      Multi-jet and \W+jets & 1900   & 500
                            &   34   &  12   \\
      \DY                   &  610   & 140
                            &    2.6 &   0.7 \\
      \hline
      Total SM background   & 3200 & 800
                            &   63 &  20     \\
      \hline
      SM+\Zprime ($M_{\Zprime}=1.5$~\si{\TeV})
      & \multicolumn{2}{c|}{-} &  130 & 28     \\
      SM+$\tilde{\nu}_{\tau}$ ($M_{\tilde{\nu}_{\tau}}=1.5$~\si{\TeV})
      & \multicolumn{2}{c|}{-} &   78 & 22     \\
      SM+QBH RS $n = 1$ ($M_{\mathrm{th}}=1.5$~\si{\TeV})
      & \multicolumn{2}{c|}{-} &   90 & 23     \\
      \hline
      Data                  & \multicolumn{2}{c|}{3239}
                            & \multicolumn{2}{c}{48}    \\
      \hline\hline
    \end{tabular}
  }
  \caption{Observed and expected numbers of (a) $e\mu$ , (b) $e\tau$, and
    (c) $\mu\tau$  events in the validation ($m_{\ell\ell^{\prime}}<600$~\si{\GeV})
    and search regions ($m_{\ell\ell^{\prime}}>600$~\si{\GeV}) for the SM
    backgrounds and the signal models considered. The quoted errors include
    statistical and systematic uncertainties. The uncertainties for the total
    background predictions account for the correlations between the
    uncertainties of the different background contributions.}
  \label{tab:yields}
\end{table}

\begin{figure}[!hbtp]
  \centering
  \subfloat[$e\mu$ channel]{
    \includegraphics[width=0.49\columnwidth]{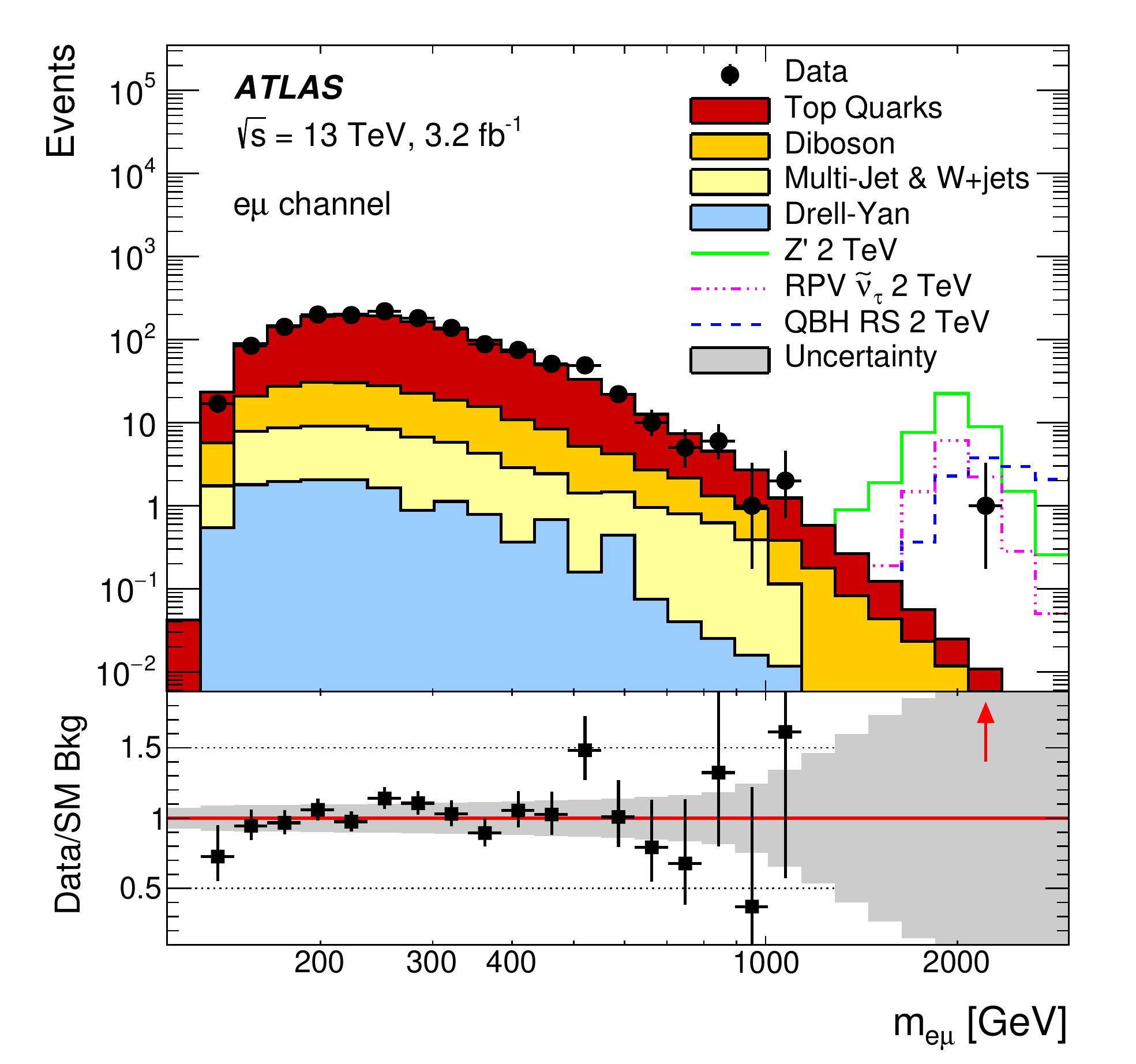}
  }
  \subfloat[$e\tau$ channel]{
    \includegraphics[width=0.49\columnwidth]{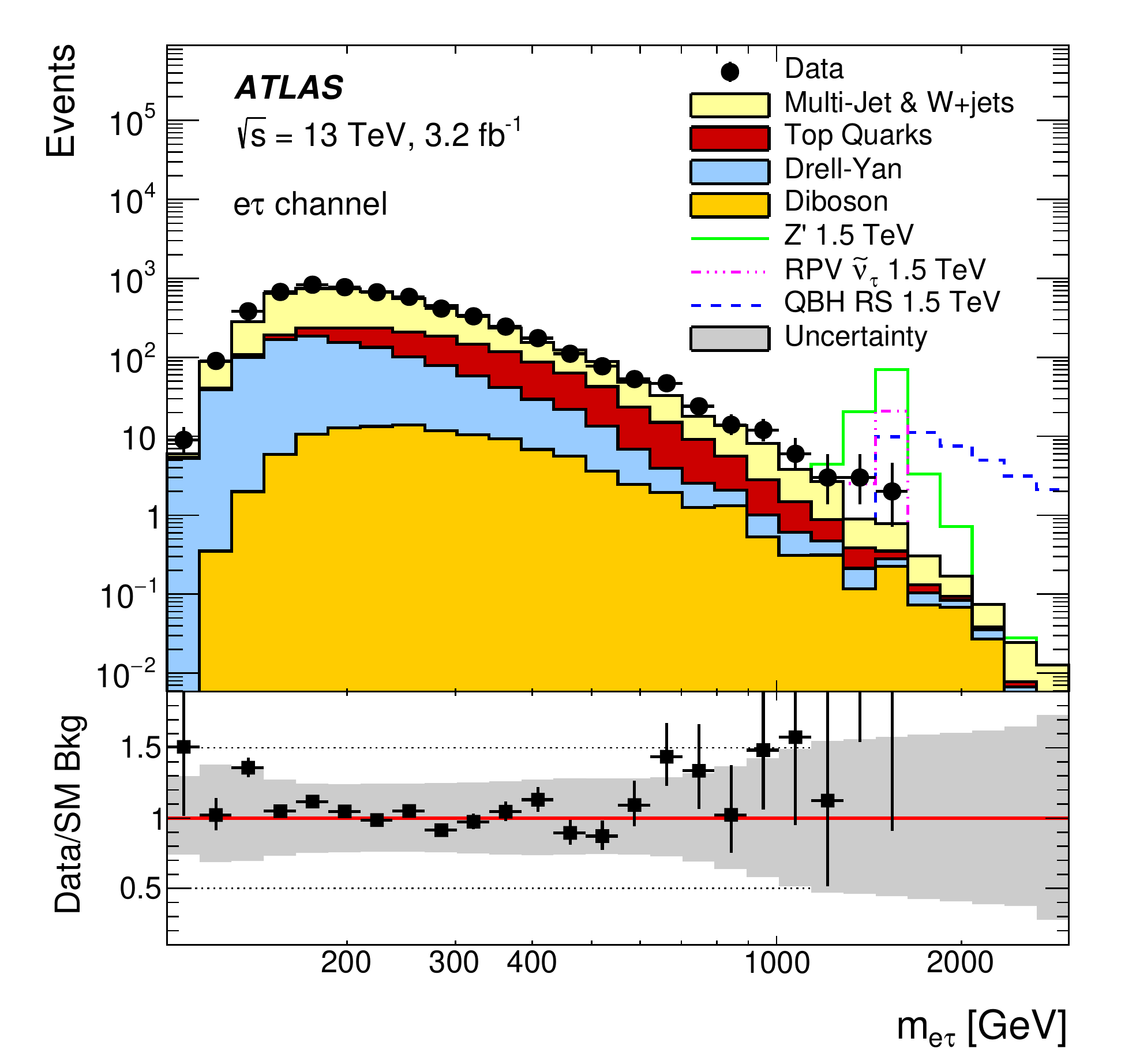}
  }\\
  \subfloat[$\mu\tau$ channel]{
    \includegraphics[width=0.49\columnwidth]{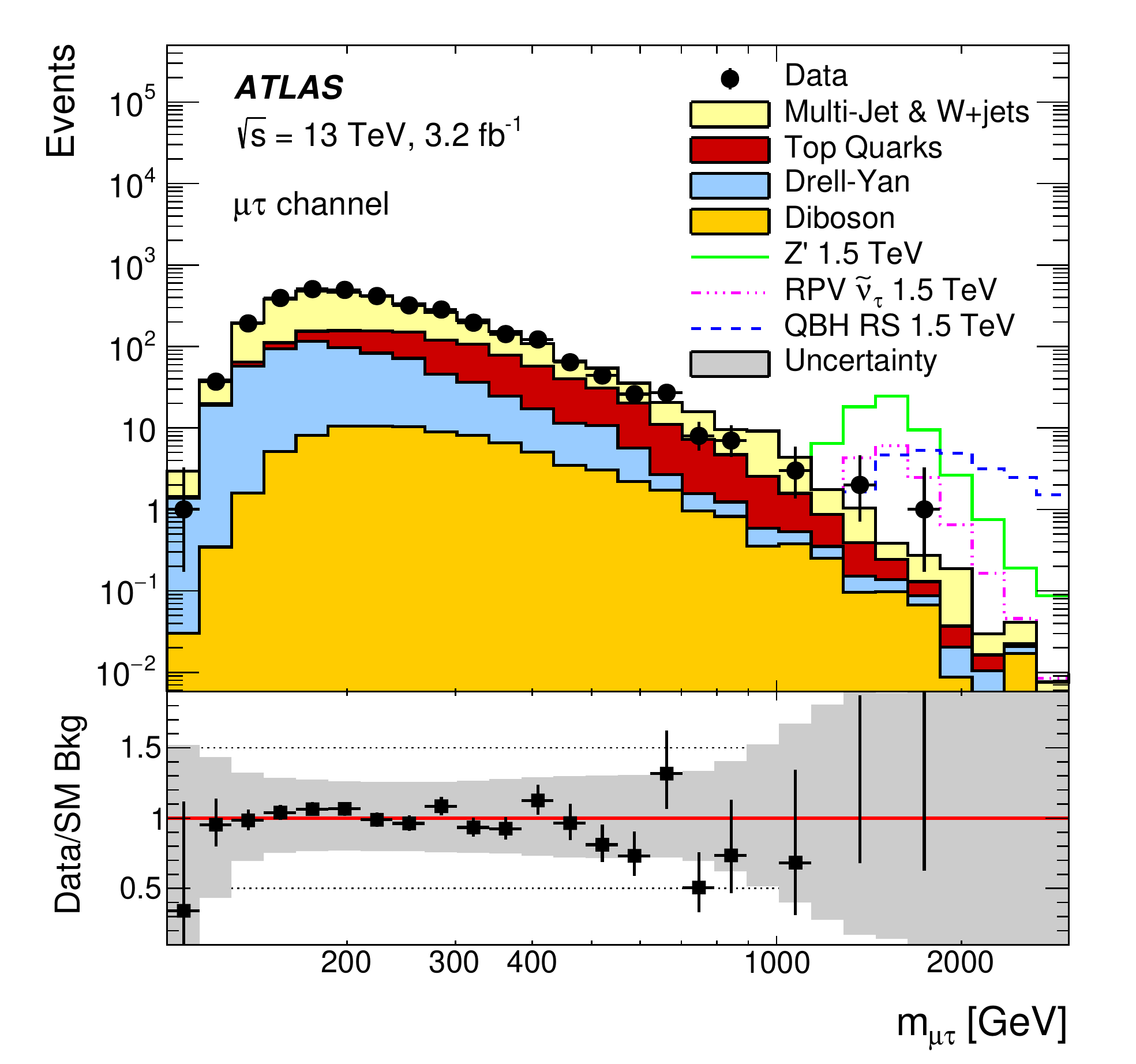}
  }
  \caption{The invariant mass distribution of final selected (a) $e\mu$,
    (b) $e\tau$ and (c) $\mu\tau$ pairs for data and MC predictions. Three
    selected signals are overlaid: a \Zprime with a mass of 2.0 and
    \SI{1.5}{\TeV}, a $\tau$ sneutrino ($\tilde{\nu}_{\tau}$) with a mass of 2.0
    and \SI{1.5}{\TeV}, and a RS quantum black hole (QBH) with a threshold mass
    of 2.0 and \SI{1.5}{\TeV}. The signal mass point shown corresponds to the highest 
    acceptance times efficiency in each channel. The error bars show the statistical uncertainty of the
    observed yields corresponding to a 68\% interval in a Poisson distribution, 
    while the band in the bottom plot includes all systematic uncertainties added in quadrature.}
  \label{fig:invmass}
\end{figure}

Figure~\ref{fig:invmass} shows the $e\mu$, $e\tau$ and $\mu\tau$ invariant mass distribution. 
The event with the largest dilepton invariant mass is found in the $e\mu$ channel 
with $m_{e\mu}$=2.1 \si{\TeV}. Since the SM expectation for $m_{e\mu} >$ 2 \si{\TeV} is 
0.02$\pm$0.02 events, the probability of observing one or more events is 2.6\%. It is then 
concluded that the observation of this high-mass candidate event is compatible with 
a statistical fluctuation and no significant excess is found over the expected background. 
Therefore, the observed data are concluded to be consistent with the SM
prediction, and model-dependent exclusion limits are extracted using the
techniques described in Section~\ref{sec:statistic}.

\begin{figure}[!htbp]
  \centering
  \subfloat[\Zprime]{
    \includegraphics[width=0.49\linewidth]{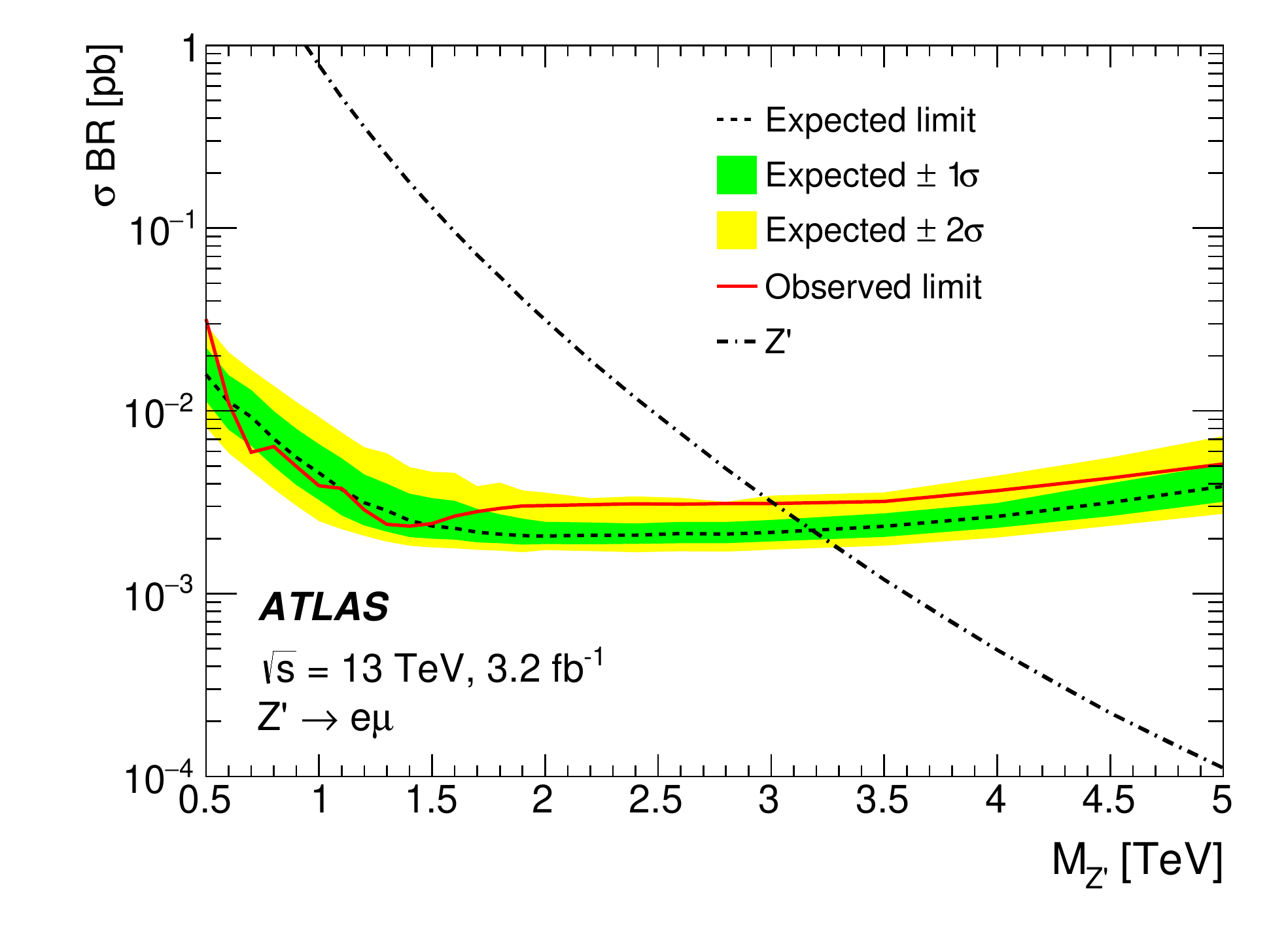}}
  \subfloat[RPV SUSY]{
    \includegraphics[width=0.49\linewidth]{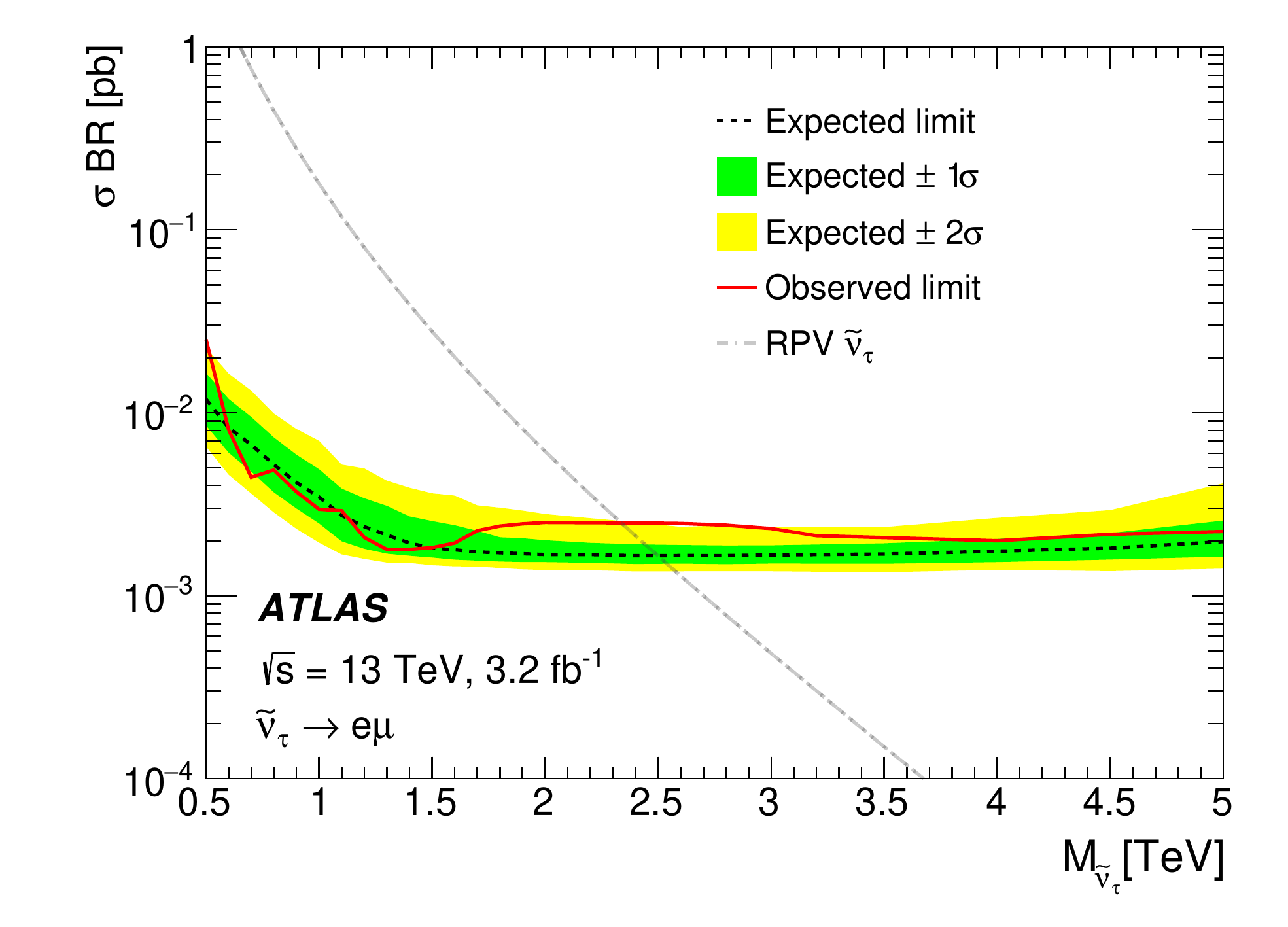}} \\
  \subfloat[QBH]{
    \includegraphics[width=0.49\linewidth]{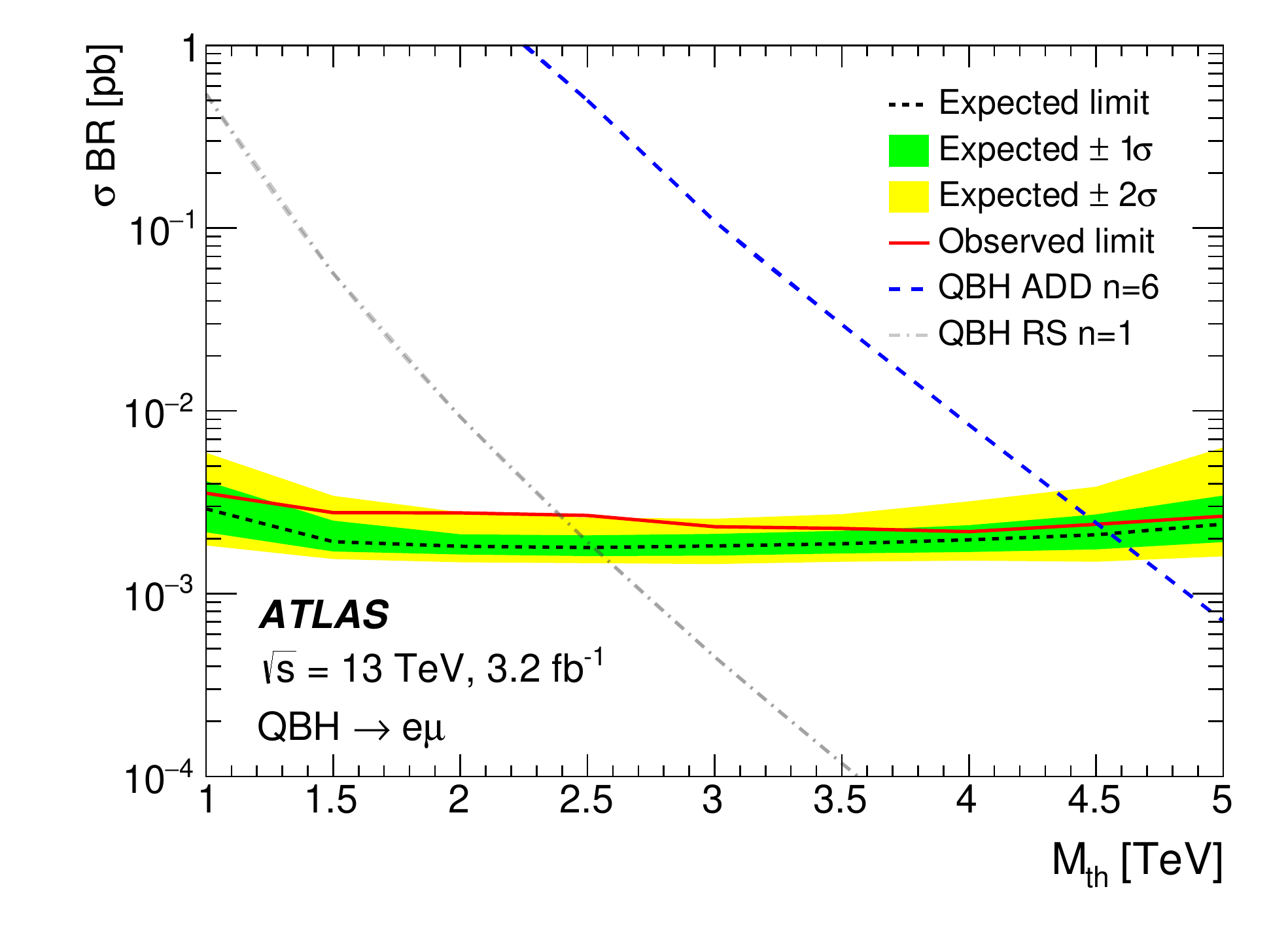}}
  \caption{The observed and expected 95\% credibility level upper limits on the
    (a) \Zprime, (b) $\tau$ sneutrino ($\tilde{\nu}_{\tau}$) and (c) QBH ADD and
    RS production cross-section times branching ratio in decays to an $e\mu$
    final state. The signal theoretical cross-section times branching ratio lines for the $Z'$ model, 
    the QBH ADD model assuming six extra dimensions and the RS model with one extra dimension 
    are obtained from the Monte Carlo generators simulating each process, 
    while the RPV SUSY $\tilde{\nu}_{\tau}$ includes the NLO $K$-factor calculated using LoopTools \cite{bib:LoopTools}.     The expected limits are plotted with the $\pm 1$
    and $\pm 2$ standard deviation uncertainty bands.}
  \label{fig:limit_emu}
\end{figure}

\begin{figure}[!htbp]
  \centering
  \subfloat[\Zprime]{
    \includegraphics[width=0.49\linewidth]{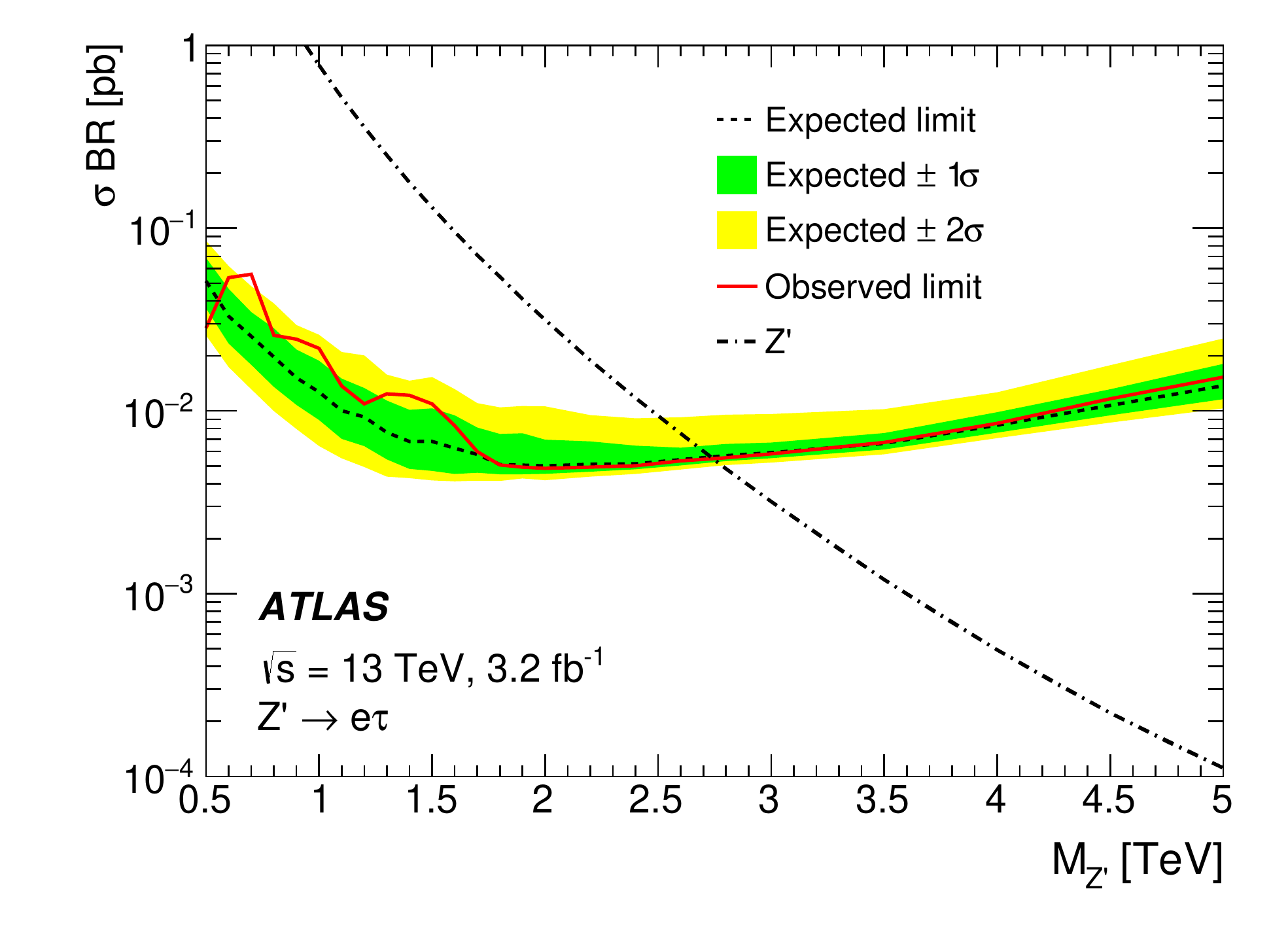}}
  \subfloat[RPV SUSY]{
    \includegraphics[width=0.49\linewidth]{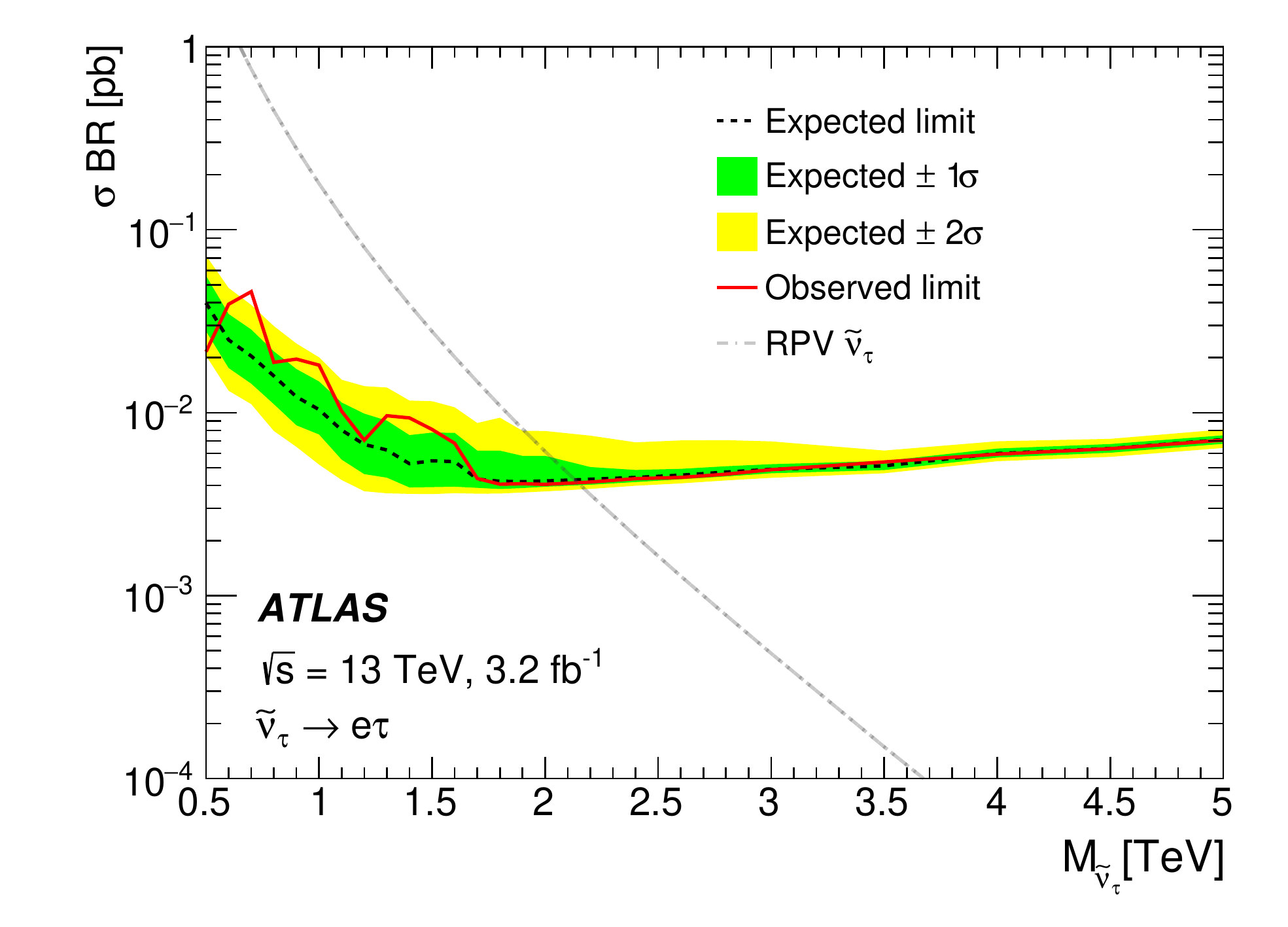}} \\
  \subfloat[QBH]{
    \includegraphics[width=0.49\linewidth]{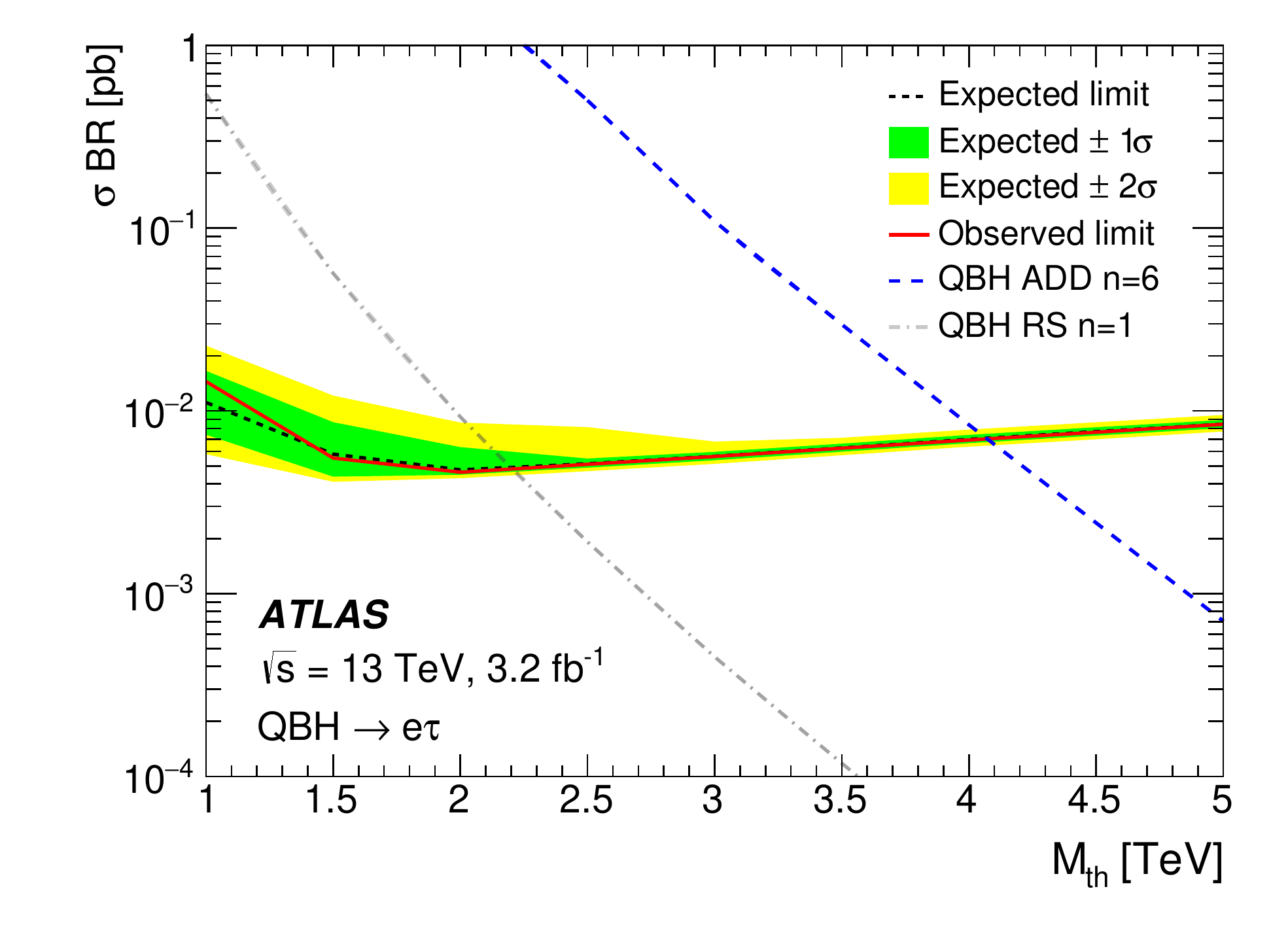}}
  \caption{The observed and expected 95\% credibility level upper limits on the
    (a) \Zprime, (b) $\tau$ sneutrino ($\tilde{\nu}_{\tau}$) and (c) QBH ADD and
    RS production cross-section times branching ratio in decays to an $e\tau$
    final state. The signal theoretical cross-section times branching ratio lines for the $Z'$ model,
    the QBH ADD model assuming six extra dimensions and the RS model with one extra dimension
    are obtained from the Monte Carlo generators simulating each process,
    while the RPV SUSY $\tilde{\nu}_{\tau}$ includes the NLO $K$-factor calculated using LoopTools \cite{bib:LoopTools}.    The expected limits are plotted with the $\pm 1$
    and $\pm 2$ standard deviation uncertainty bands.}
  \label{fig:limit_etau}
\end{figure}

\begin{figure}[!htbp]
  \centering
  \subfloat[\Zprime]{
    \includegraphics[width=0.49\linewidth]{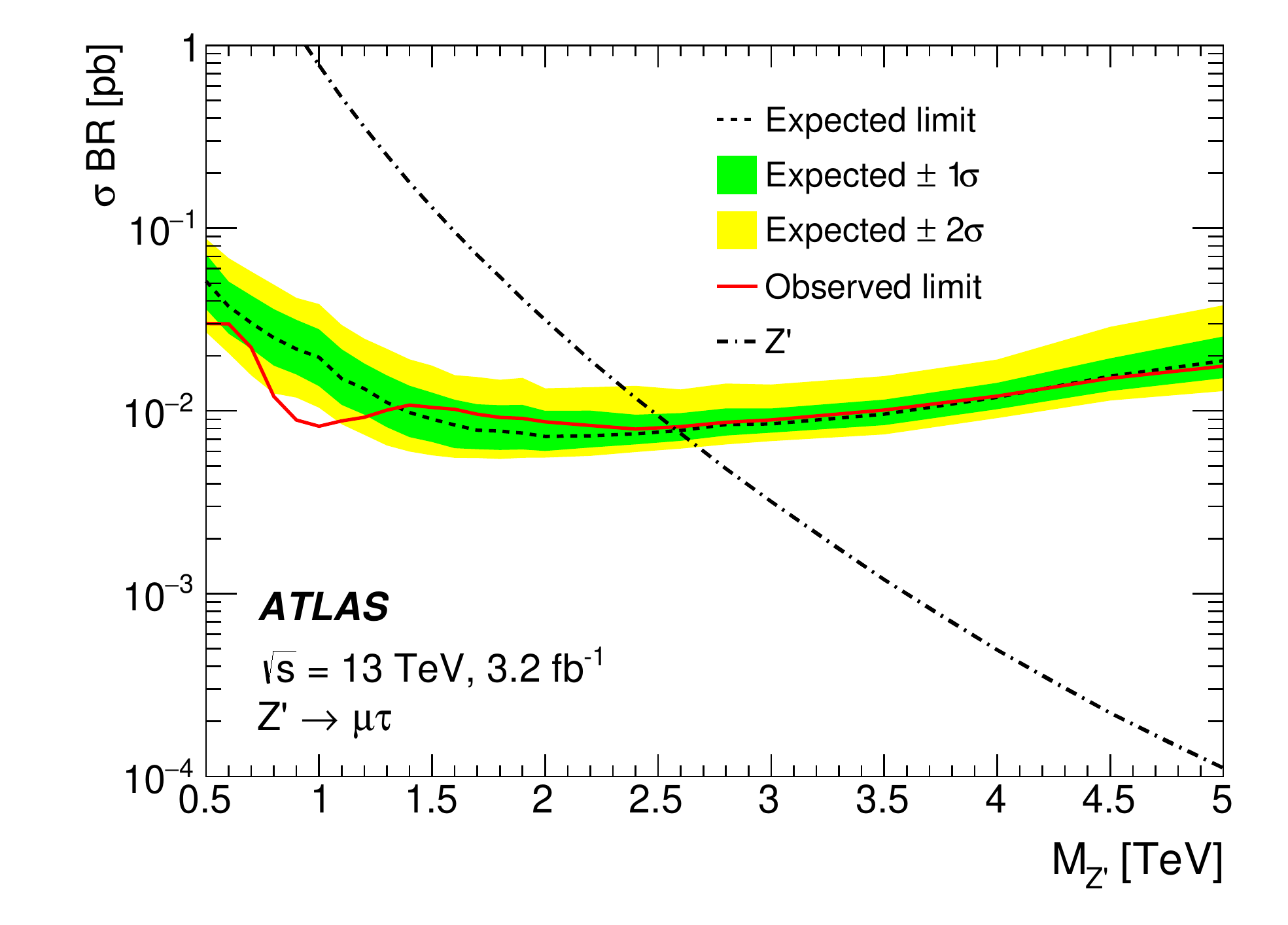}}
  \subfloat[RPV SUSY]{
    \includegraphics[width=0.49\linewidth]{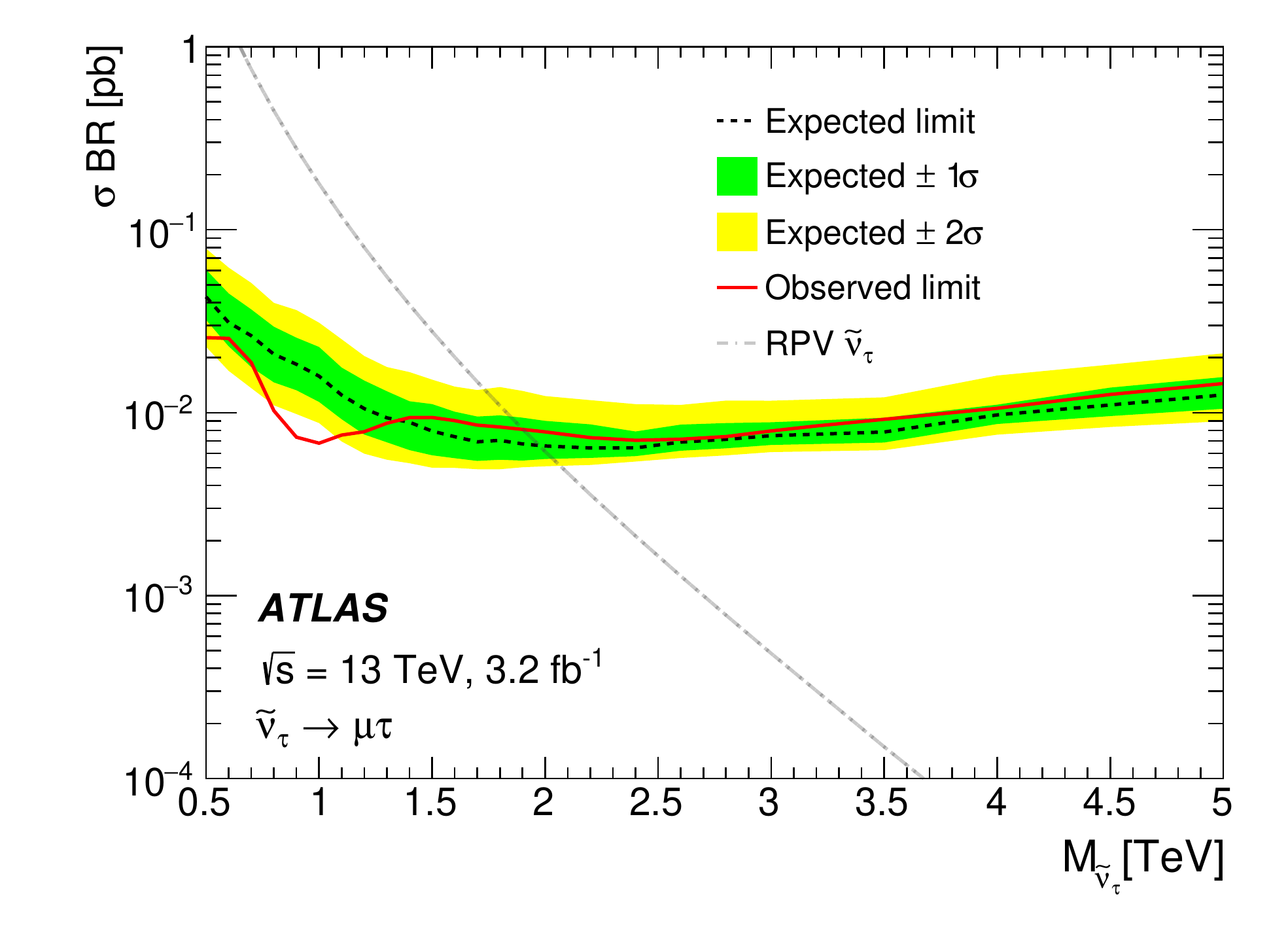}} \\
  \subfloat[QBH]{
    \includegraphics[width=0.49\linewidth]{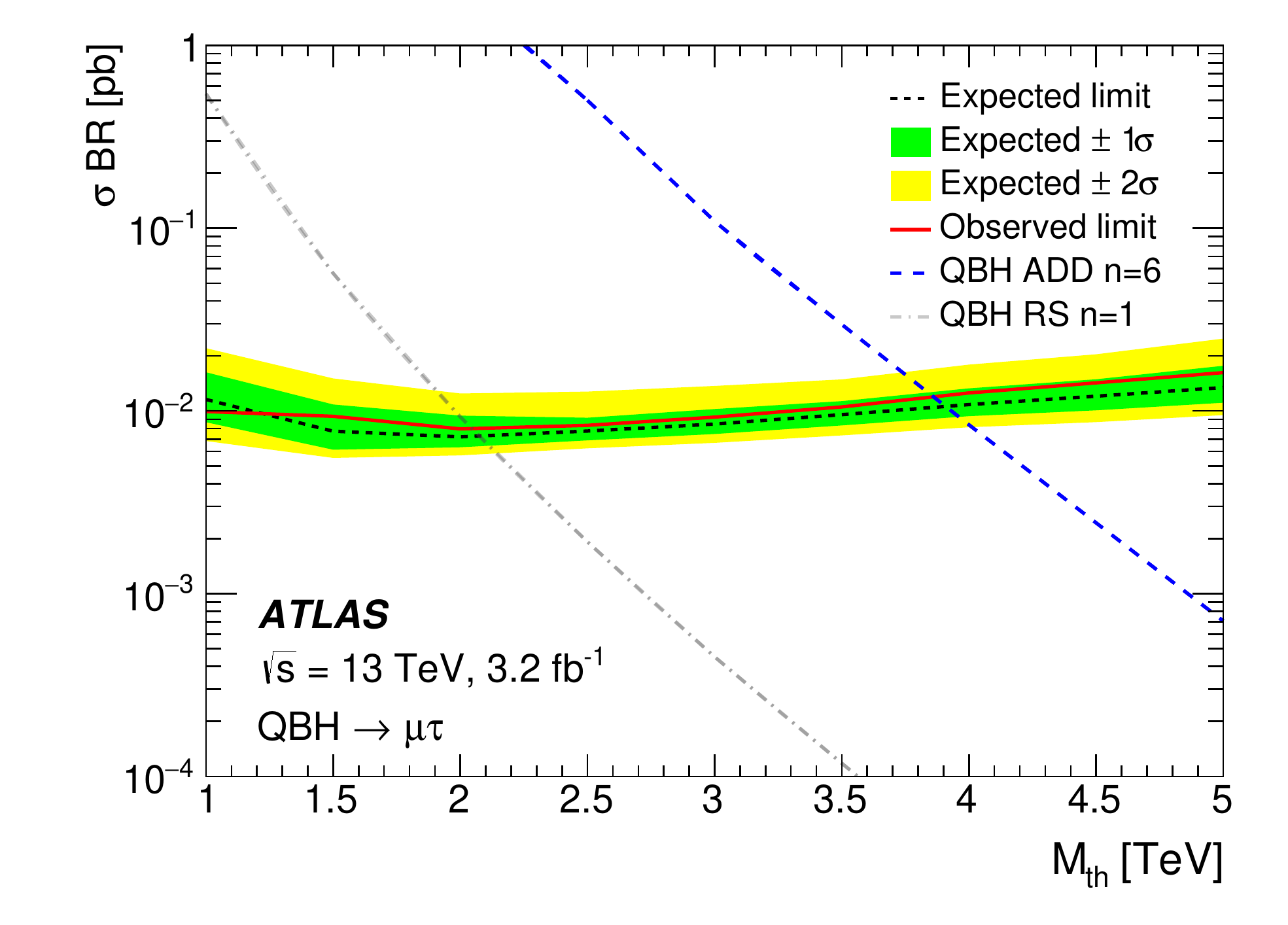}} 
  \caption{The observed and expected 95\% credibility level upper limits on the
    (a) \Zprime, (b) $\tau$ sneutrino ($\tilde{\nu}_{\tau}$) and (c) QBH ADD and
    RS production cross-section times branching ratio in decays to an $\mu\tau$
    final state. The signal theoretical cross-section times branching ratio lines for the $Z'$ model,
    the QBH ADD model assuming six extra dimensions and the RS model with one extra dimension
    are obtained from the Monte Carlo generators simulating each process,
    while the RPV SUSY $\tilde{\nu}_{\tau}$ includes the NLO $K$-factor calculated using LoopTools \cite{bib:LoopTools}.    The expected limits are plotted with
    the $\pm 1$ and $\pm 2$ standard deviation uncertainty bands.}
  \label{fig:limit_mutau}
\end{figure}

Figures~\ref{fig:limit_emu}, \ref{fig:limit_etau} and \ref{fig:limit_mutau}
show the 95\% CL expected and observed upper limits on the production
cross-section times branching ratio of the \Zprime, RPV SUSY $\tilde{\nu}_{\tau}$ and QBH models for each
of the final states considered. The extracted limits worsen for signal masses
above 2.5 (1.5) \si{\TeV} in the $e\mu$ ($e\tau$ and $\mu\tau$) channel due to
a decrease in the lepton reconstruction efficiency at very high $p_{\mathrm{T}}$.
Results are summarised in Table~\ref{tab:limits}. The $A\epsilon$ of the ADD
and RS QBH models were found to agree within 1\% and therefore the same curve
is used for the limit extraction.

\begin{table}[!htbp]
  \centering
  \begin{tabular}{l|ccc|ccc}
    \hline\hline
    \multirow{2}{*}{Model}
    & \multicolumn{3}{c|}{Expected Limit [\si{\TeV}]}
    & \multicolumn{3}{c}{Observed Limit [\si{\TeV}]} \\
    \cline{2-7}
    & \;\;$e\mu$ & \;\;$e\tau$ & $\mu\tau$
    & \;\;$e\mu$ & \;\;$e\tau$ & $\mu\tau$ \\
    \hline
    \Zprime
    & \;\;3.2 & \;\;2.7 & 2.6 & \;\;3.0 & \;\;2.7 & 2.6 \\
    RPV SUSY $\tilde{\nu}_{\tau}$
    & \;\;2.5 & \;\;2.1 & 2.0 & \;\;2.3 & \;\;2.2 & 1.9 \\
    QBH ADD $n=6$
    & \;\;4.6 & \;\;4.1 & 3.9 & \;\;4.5 & \;\;4.1 & 3.9 \\
    QBH RS $n=1$
    & \;\;2.5 & \;\;2.2 & 2.1 & \;\;2.4 & \;\;2.2 & 2.1 \\
    \hline\hline
  \end{tabular}
  \caption{Expected and observed 95\% credibility level lower limits on the
    mass of a \Zprime with lepton-flavour-violating couplings, a supersymmetric
    $\tau$ sneutrino ($\tilde{\nu}_{\tau}$) with $R$-parity-violating couplings,
    and the threshold mass for quantum black hole production for the ADD $n=6$
    and RS $n=1$ models. Limits for all channels are reported.}
  \label{tab:limits}
\end{table}

\FloatBarrier

\section{Conclusions}
\label{sec:conclusion}
A search for a heavy particle decaying into an $e\mu$, $e\tau$ or $\mu\tau$
$(\ell\ell^{\prime})$ final state is conducted, using \SI{3.2}{\per\femto\barn}
of $\sqrt{s}=13$~\si{\TeV} proton--proton collision data recorded by the ATLAS
detector at the Large Hadron Collider.
The data are found to be consistent with the Standard Model prediction
in both the validation region ($m_{\ell\ell^{\prime}}<600$~\si{\GeV}) and search
region ($m_{\ell\ell^{\prime}}>600$~\si{\GeV}). With no evidence of new physics,
Bayesian lower limits at 95\% credibility level are set on the mass of a
\Zprime vector boson with lepton-flavour-violating couplings at 3.0, 2.7 and
\SI{2.6}{\TeV} separately for $e\mu$, $e\tau$ and $\mu\tau$ pairs, and a
supersymmetric $\tau$ sneutrino ($\tilde{\nu}_{\tau}$) with $R$-parity-violating
couplings at 2.3, 2.2 and \SI{1.9}{\TeV}. The results are also interpreted as limits on the
threshold mass for quantum black hole production. The exclusion limits extracted on the mass of a 
\Zprime and the supersymmetric $\tau$ sneutrino extend by around 20\% those reported by ATLAS and CMS 
using the full dataset at $\sqrt{s}$ = 8~\si{\TeV}.

\clearpage

\section*{Acknowledgements}

% Acknowledgements for papers with collision data
% Version 2-Aug-2016

% Standard acknowledgements start here
%----------------------------------------------
We thank CERN for the very successful operation of the LHC, as well as the
support staff from our institutions without whom ATLAS could not be
operated efficiently.

We acknowledge the support of ANPCyT, Argentina; YerPhI, Armenia; ARC, Australia; BMWFW and FWF, Austria; ANAS, Azerbaijan; SSTC, Belarus; CNPq and FAPESP, Brazil; NSERC, NRC and CFI, Canada; CERN; CONICYT, Chile; CAS, MOST and NSFC, China; COLCIENCIAS, Colombia; MSMT CR, MPO CR and VSC CR, Czech Republic; DNRF and DNSRC, Denmark; IN2P3-CNRS, CEA-DSM/IRFU, France; GNSF, Georgia; BMBF, HGF, and MPG, Germany; GSRT, Greece; RGC, Hong Kong SAR, China; ISF, I-CORE and Benoziyo Center, Israel; INFN, Italy; MEXT and JSPS, Japan; CNRST, Morocco; FOM and NWO, Netherlands; RCN, Norway; MNiSW and NCN, Poland; FCT, Portugal; MNE/IFA, Romania; MES of Russia and NRC KI, Russian Federation; JINR; MESTD, Serbia; MSSR, Slovakia; ARRS and MIZ\v{S}, Slovenia; DST/NRF, South Africa; MINECO, Spain; SRC and Wallenberg Foundation, Sweden; SERI, SNSF and Cantons of Bern and Geneva, Switzerland; MOST, Taiwan; TAEK, Turkey; STFC, United Kingdom; DOE and NSF, United States of America. In addition, individual groups and members have received support from BCKDF, the Canada Council, CANARIE, CRC, Compute Canada, FQRNT, and the Ontario Innovation Trust, Canada; EPLANET, ERC, FP7, Horizon 2020 and Marie Sk{\l}odowska-Curie Actions, European Union; Investissements d'Avenir Labex and Idex, ANR, R{\'e}gion Auvergne and Fondation Partager le Savoir, France; DFG and AvH Foundation, Germany; Herakleitos, Thales and Aristeia programmes co-financed by EU-ESF and the Greek NSRF; BSF, GIF and Minerva, Israel; BRF, Norway; Generalitat de Catalunya, Generalitat Valenciana, Spain; the Royal Society and Leverhulme Trust, United Kingdom.

The crucial computing support from all WLCG partners is acknowledged gratefully, in particular from CERN, the ATLAS Tier-1 facilities at TRIUMF (Canada), NDGF (Denmark, Norway, Sweden), CC-IN2P3 (France), KIT/GridKA (Germany), INFN-CNAF (Italy), NL-T1 (Netherlands), PIC (Spain), ASGC (Taiwan), RAL (UK) and BNL (USA), the Tier-2 facilities worldwide and large non-WLCG resource providers. Major contributors of computing resources are listed in Ref.~\cite{ATL-GEN-PUB-2016-002}.
%----------------------------------------------

\clearpage
\printbibliography
\clearpage

% ATLAS Collaboration author list
% Data extracted on 27-Jul-2016 for paper reference EXOT-2015-20
%\documentclass[11pt]{article}
%\usepackage{a4wide}%\begin{document}
\begin{flushleft}
{\Large The ATLAS Collaboration}

\bigskip

M.~Aaboud$^\textrm{\scriptsize 136d}$,
G.~Aad$^\textrm{\scriptsize 87}$,
B.~Abbott$^\textrm{\scriptsize 114}$,
J.~Abdallah$^\textrm{\scriptsize 65}$,
O.~Abdinov$^\textrm{\scriptsize 12}$,
B.~Abeloos$^\textrm{\scriptsize 118}$,
R.~Aben$^\textrm{\scriptsize 108}$,
O.S.~AbouZeid$^\textrm{\scriptsize 138}$,
N.L.~Abraham$^\textrm{\scriptsize 152}$,
H.~Abramowicz$^\textrm{\scriptsize 156}$
% ...
\bigskip
\\
% ...
$^{12}$ Institute of Physics, Azerbaijan Academy of Sciences, Baku, Azerbaijan\\
% ...
$^{65}$ University of Iowa, Iowa City IA, United States of America\\
% ...
$^{87}$ CPPM, Aix-Marseille Universit{\'e} and CNRS/IN2P3, Marseille, France\\
% ...
$^{108}$ Nikhef National Institute for Subatomic Physics and University of Amsterdam, Amsterdam, Netherlands\\
% ...
$^{114}$ Homer L. Dodge Department of Physics and Astronomy, University of Oklahoma, Norman OK, United States of America\\
% ...
$^{118}$ LAL, Univ. Paris-Sud, CNRS/IN2P3, Universit{\'e} Paris-Saclay, Orsay, France\\
% ...
$^{136}$ $^{(a)}$ Facult{\'e} des Sciences Ain Chock, R{\'e}seau Universitaire de Physique des Hautes Energies - Universit{\'e} Hassan II, Casablanca; $^{(b)}$ Centre National de l'Energie des Sciences Techniques Nucleaires, Rabat; $^{(c)}$ Facult{\'e} des Sciences Semlalia, Universit{\'e} Cadi Ayyad, LPHEA-Marrakech; $^{(d)}$ Facult{\'e} des Sciences, Universit{\'e} Mohamed Premier and LPTPM, Oujda; $^{(e)}$ Facult{\'e} des sciences, Universit{\'e} Mohammed V, Rabat, Morocco\\
% ...
$^{138}$ Santa Cruz Institute for Particle Physics, University of California Santa Cruz, Santa Cruz CA, United States of America\\
% ...
$^{152}$ Department of Physics and Astronomy, University of Sussex, Brighton, United Kingdom\\
% ...
$^{156}$ Raymond and Beverly Sackler School of Physics and Astronomy, Tel Aviv University, Tel Aviv, Israel\\
% ...
$^{a}$ Also at Department of Physics, King's College London, London, United Kingdom\\
$^{b}$ Also at Institute of Physics, Azerbaijan Academy of Sciences, Baku, Azerbaijan\\
$^{c}$ Also at Novosibirsk State University, Novosibirsk, Russia\\
$^{d}$ Also at TRIUMF, Vancouver BC, Canada\\
$^{e}$ Also at Department of Physics {\&} Astronomy, University of Louisville, Louisville, KY, United States of America\\
% ...
\end{flushleft}

%\end{document}
% Created with xml2latex.py

\clearpage

\end{document}